\documentclass[a4paper,11pt]{article}
\usepackage{epsf}
\newcommand{\bmat}{\left(\begin{array}}
\newcommand{\emat}{\end{array}\right)}

\def\t{\times}

\def\beq{\begin{equation}}
\def\eeq{\end{equation}}
\def\beqa{\begin{eqnarray}}
\def\eeqa{\end{eqnarray}}
\def\t{\times}

\def\-{\hphantom{-}}

\def\s2{\frac{1}{\sqrt2}}

\def\beq{\begin{equation}}
\def\eeq{\end{equation}}
\def\beqa{\begin{eqnarray}}
\def\eeqa{\end{eqnarray}}
\def\ba{\begin{array}}
\def\ea{\end{array}}

\def\IF{\relax{\rm I\kern-.18em F}}
\def\II{\relax{\rm I\kern-.18em I}}
\def\IP{\relax{\rm I\kern-.18em P}}
\def\IC{\relax\hbox{\kern.25em$\inbar\kern-.3em{\rm C}}}
\def\IR{\relax{\rm I\kern-.18em R}}

\def\cp{{\cal P}}

\def\Dsl{\,\raise.15ex\hbox{/}\mkern-13.5mu D} %this one can be subscripted
\def\IZ{Z\kern-.4em  Z}

 \def\cp#1{\relax\ifmmode {\IP\kern-2pt{}_{#1}}\else $\IP\kern-2pt{}_{#1}$\=fi}
\catcode`\@=11
\newdimen\@rotdimen
\newbox\@rotbox

\def\@vspec#1{\special{ps:#1}}%  passes #1 verbatim to the output
\def\@rotstart#1{\@vspec{gsave currentpoint currentpoint translate
   #1 neg exch neg exch translate}}% #1 can be any origin-fixing transformation
\def\@rotfinish{\@vspec{currentpoint grestore moveto}}% gets back in synch
\def\@rotr#1{\@rotdimen=\ht#1\advance\@rotdimen by\dp#1%
   \hbox to\@rotdimen{\hskip\ht#1\vbox to\wd#1{\@rotstart{90 rotate}%
   \box#1\vss}\hss}\@rotfinish}
\def\@rotl#1{\@rotdimen=\ht#1\advance\@rotdimen by\dp#1%
   \hbox to\@rotdimen{\vbox to\wd#1{\vskip\wd#1\@rotstart{270 rotate}%
   \box#1\vss}\hss}\@rotfinish}%
\def\@rotu#1{\@rotdimen=\ht#1\advance\@rotdimen by\dp#1%
   \hbox to\wd#1{\hskip\wd#1\vbox to\@rotdimen{\vskip\@rotdimen
   \@rotstart{-1 dup scale}\box#1\vss}\hss}\@rotfinish}%
\def\@rotf#1{\hbox to\wd#1{\hskip\wd#1\@rotstart{-1 1 scale}%
   \box#1\hss}\@rotfinish}%
\def\rotate{\@ifnextchar[{\@rotate}{\@rotate[l]}}
\def\@rotate[#1]#2{\setbox\@rotbox=\hbox{#2}\@nameuse{@rot#1}\@rotbox}

\catcode`\@=12
\begin{document}

%----------------------------------------------------------------------%
%  numbering equations with section number
%----------------------------------------------------------------------%
\makeatletter \@addtoreset{equation}{section} \makeatother
\renewcommand{\theequation}{\thesection.\arabic{equation}}
%----------------------------------------------------------------------%
%  title page
%----------------------------------------------------------------------%
\pagestyle{empty}
%----------------------------------------------------------------------%
%  Resetting of counters
%----------------------------------------------------------------------%
%\setcounter{page}{0}
\pagestyle{empty}
%\vspace{0.5in}
%\rightline{FTUAM-03-09}
%\rightline{IFT-UAM/CSIC-03-17}
\rightline{\today}
\vspace{3.0cm}
\setcounter{footnote}{0}

%\begin{center}
%\underline{\LARGE{****  DRAFT VERSION *****}}
%\end{center}

\begin{center}
\LARGE{\bf Euler Top Dynamics of Nambu-Goto P-Branes}
\\[8mm]
%\medskip
{\Large{ Minos Axenides$^{1}$ and Emmanuel Floratos$^{1,2}$}}
% and Christos Kokorelis$^{1}$}}
\\[6mm]
 \normalsize{\em $^1$ Institute of Nuclear Physics, N.C.S.R. Demokritos,
GR-15310, Athens, Greece}\\
\normalsize{\em $^2$  Department of Physics, Univ. of Athens,
GR-15771 Athens, Greece}\\[5mm]  axenides@inp.demokritos.gr \ \ \   mflorato@phys.uoa.gr 
\end{center}
\vspace{6.0mm}
%%%%%%%%%%%%%%%%%%%%%%%%%%%%%%%%%%%%%%%
%\begin{center}
%\begin{minipage}[h]{14.5cm}
\begin{center}
{\large {\bf Abstract}}
\end{center}
%We find new two and three dimensional exact spinning solutions of the M(atrix) Model and 
%p-Branes in Light-Cone Flat space-times with spherical and toroidal compactifications. They minimize the Energy functional
%as higher dimensional Euler-Tops, they are isometrically embedable in certain gravitational backgrounds
%with or without fluxes. They represent compact non-perturbative solitonic states of type IIA-B superstrings
%in Higher-Dimensional p-p waves. Their energy scales as $\sim 1/g_{IIA-B}$ with respect
%to the string coupling constant and depends linearly on the anugular monenta.
\vspace{5.0mm}
We propose a method to obtain new exact solutions of spinning p-branes in flat space-times 
for any p , which manifest themselves as higher dimensional Euler Tops and minimize their
energy functional. We provide concrete examples for the case of spherical topology
$S^{2}, S^{3}$ and rotational symmetry $ \prod_{i}SO(q_{i})$.
In the case of toroidal topology $ T^{2}, T^{3} $ the rotational symmetry is 
$ \prod_{i} SU(q_{i}) $ with  m target dimensions being
compactified on the torus $ T^{m}$ . By double dimensional reduction the Light Cone
Hamiltonians of $ T^{2}, T^{3} $ reduce to those of closed string $ S^{1} $ and $ T^{2} $
membranes respectively. The solutions are interpreted as non-perturbative spinning soliton states
of type $ IIA-IIB$ superstrings.

%We find new exact spinning solutions of  p-Branes for $p=2,3$ in Light Cone flat
%spacetimes with spherical($S^{2},S^{3}$)  and Toroidal Topology ($T^{2},T^{3}$) 
%which manifest as solutions of the M(atrix) Model ($ S^{2},T^{2} $).
%They minimize the energy as higher dimensional Euler Tops and they are isometrically embedable 
%in certain gravitational backgrounds. As such they represent non-perurbative spinning states
%of type IIA-B superstrings.  

\newpage

%----------------------------------------------------------------------%
%  Resetting of counters
%----------------------------------------------------------------------%
\setcounter{page}{1} \pagestyle{plain}
\renewcommand{\thefootnote}{\arabic{footnote}}
\setcounter{footnote}{0}
%----------------------------------------------------------------------%
%  Paper begins
%----------------------------------------------------------------------%
\tableofcontents
\section{Introduction}

One of the most important discoveries in theoretical physics in the last few years has been the
connection of the strongly coupled gauge theories to perturbative gravity through the Maldacena conjecture 
\cite{Mal}. This is only one spectacular result of the UV/IR relation and Holography, 
discovered in non-commutative geometry of D-branes in gravitational backgrounds with fluxes \cite{Sushoof}. 
In order to understand this connection, the
most important tool  has been the comparison of the energy spectra
of rotating strings, D-branes, p-branes and/or even matrix model rotating solutions in various gravitational 
backgrounds
with the anomalous dimensions of composite operators of the boundary gauge, or more generally, of the conformal field theory.
More recently such a comparison has been in the focus due to their connection with Bethe ansatz methods of obtaining
the spectra of integrable spin chain models. Impressive agreement on both sides has been obtained \cite{Plef}. 

Another interesting development has come about by the use of rotating $D_{3}$ branes  in the presence of fluxes giving rise to a stringy exclusion principle as well as the 
notion of giant graviton\cite{Mal, GSToum}. Rotating solutions in backgrounds of pp waves along with their dielectric 
behaviour in the presence of fluxes has been studied. Their connection with the BPS sector of $N=4$ Super-Yang-Mills theories has 
been established\cite{Jabb}. In a completely different direction Matrix or brane solutions have been interpreted in the framework of
Matrix Cosmology\cite{FGS}. An important class of new nonrelativistic Newton-Hook cosmologies
appears from deSitter spacetime backgrounds in the Newton-Hooke limit of
$\frac{\Lambda c^2}{3}=$constant as $\Lambda \rightarrow 0$ and 
$c^2\rightarrow\infty $ \cite{Gibb}.

Rotating Solutions for strings and p-branes were studied in the first few years of the 
development of this 
field by searching for massless particles in their spectra\cite{CoTuck}. In the case of superstring theory the full supergravity
multiplets have been discovered raising, as a consequence, the string to the status of a more fundamental theory. 
Much later it was understood that
other extended objects, such as D-branes \cite{Polc} are connected through nonperturbative dualities. This has led to the creation of the hypothesis of
M-theory and Matrix model\cite {Town}.

It is obvious from the above that there is a strong motivation for a more exhaustive search for non-perturbative soliton solutions of string
theories such as membranes , 3-branes and/or matrix model solutions in various backgrounds 
with or without fluxes. All of these should be compared with known spectra of operators of gauge or conformal field theories.
Another interesting application can be the determination of the quantum effective Hamiltonian for p-branes as
fundamental objects\cite{Mets}.

In this paper we propose a method to extract new solutions for spinning p-branes in the Light Cone
spacetime for any p. By providing concrete examples we continue our search for membrane or matrix solutions 
\cite{AFP} in a more systematic
way, thus exhausting the class of rotating solutions for $S^{2}$, $T^{2}$ in flat space times with  
toroidal compactifications which are consistent with our method. We demonstrate  that
the rigid body type of Eulerian motion minimizes the energy with a given conserved angular momentum. 
We extend these solutions
to higher dimensionalities of the extended object (e.g. $p=3$ for $S^{3}, \ T^{3}$). The method can be applied to any p . 
In order to
achieve this we make use of the lightcone gauge where Nambu brackets play a natural role by expressing the 
extension of the 
infinite gauge group from area preserving diffeomorphisms ($p=2$) to p-volume preserving diffeomorphisms.

Although for $p\geq 3$ there have been efforts to formulate corresponding matrix models 
\cite{Ramg} we will not attempt to apply our method to these models.
We believe that if fluxes are not present ( absence of Dielectric Myers effect)
matrix models fuzzify only membranes ($p=2$) because of the generic two discrete indices 
of matrices. Higher values of p which constitute generalizations to multiindexed matrices
 with p discrete indices are necessary. 
Unknown mathematical structures for multiplication and more general algebraic operations of these
objects must be sought for. 

We will restrict ourselves to flat backgrounds with toroidal compactifications. 
We observe that the world volumes of our solutions live in submanifolds with spherical or toroidal geometry.
This property may possibly be used to embed isometrically 
our solutions into curved spacetime backgrounds with the same world volumes as minimal 
submanifolds. These embeddings might provide solutions of the extended objects in these 
specific curved backgrounds (e.g. $AdS^5 \t  S^5$,$AdS^7 \t S^4$ and $G^2$ )\cite{Alish}. 

We organize our work as follows: 

In ch.2 we write down the equations of motion and their constraints in the lightcone gauge
for p-branes and the matrix model in flat spacetimes. We introduce the Nambu brackets, a
minimun set of their properties as well as the definitions of their p-volume preserving diffeomorphisms. 

In ch.3 we
construct the extension of the Euler Top equations of motion to higher dimensions which are appropriate for p-branes.
We write the relation between their total energy , angular momenta and generalized angular velocities. 
We provide the
NASCs in order that a p-brane solution can be characterized as higher dim. Euler Top ("P-Branetops"). 

In ch.4 we apply the Euler Top
formalism in order to present  solutions for spinning $S^2$ and $S^3$ branes with rotational symmetries $ \prod_{i} SO(q_{i})$. 

In ch.5 we examine the case of the 
spinning toroidal 
$T^{2}$ and $T^{3}$ branes including toroidal compactifications with rotational symmetries
$ \prod_{i} SU(q_{i})$.

In the conclusions we interpret the solutions as nonperturbative type IIA-B solitons.
Their energy is related 
non-perturbatively to the corresponding string coupling constants \cite{Russo}. We close by
discussing  the relevance of our results to 
other recent 
work in the literature.

\newpage

\section{Lightcone Equations of Motion for P-Branes and  \\ Nambu Brackets.}

The Light Cone gauge of Nambu-Goto p-branes 
for flat space-times has been worked out in detail two decades ago \cite{Berg}. 
The resulting Hamiltonian for the bosonic sector with zero flux background 
is given by :

\beq
H = \frac{T_{p}}{2} \int d^{p}\xi  \sqrt{\gamma} \ \ 
[ \dot{X}^{{i}^2} \ + det \ 
[\partial_{\alpha } X^i \partial_{\beta} X^i  ] ], \ \ \ \ i=1,\ldots,D-2  \ \ \ \alpha,
\beta = 1,\ldots ,p 
\eeq
$T_{p}$ is the brane tension, $d^{p}\xi \sqrt {\gamma}$ is the volume element in $\xi$-space 
\beq
 \partial _{\alpha} = \frac{\partial}{\partial \xi^{\alpha}} \ \ \alpha=1,\ldots,p 
\eeq
It is easy to observe that the potential energy term of the Hamiltonian 
can be rewritten in terms of the Nambu p-bracket .

\beq
det \ [\partial_{\alpha}X_{i} \partial_{\beta}X_{i} ] = \frac{1}{p!} \ \ \ \sum_{ i_{1},\ldots, i_{p}=1}^{D-2}
\{ X_{i_{1}},\ldots, X_{i_{p}}\}^{2}
\eeq
where
\beq
\{ f_{1},\ldots,f_{p}\} \equiv  \frac{1}{\sqrt{\gamma}} \epsilon^{\alpha_{1}\cdots \alpha_{p}} \partial_
{\alpha_{1}}f_{1}\cdots\partial_{\alpha_{p}}f_{p}, \ \ \ \ \ \alpha_{1},\ldots, \alpha_{p}=1,\ldots, p
\eeq
The eqs of motion in terms of Nambu p-brackets read:

\beqa
\ddot{X} \ &=& \ \frac{1}{(p-1)\ !} \{ \{ X_i, X_{j_{1}},\ldots, X_{j_{D-1}} \} , X_{j_{1}}, \ldots, X_{j_{p-1}} \} 
  \\ & &
  \ \ \ \ \ \ \ \ \ \ \ \ \ \ \ \ \ \ \ i,j_{1},\ldots, j_{p-1}= 1,2,\ldots, D-2 \nonumber 
\eeqa
The p-dimensional reparametrization invariance of the Lagrangian has been reduced after LC gauge fixing
to p-volume preserving diffeomorphisms of the brane manifold $M_{p}$ , $VolDiffs [M_{p}]$ \cite{Berg}. This infinite 
dimensional gauge group contains elements not connected with the identity depending on the topology 
of $M_{p}$. The $ Vol Diffs [M_{p}] $ connected to the identity gauge transformations are generated 
by the constraints 
 
\beqa
\{  \dot{X}_i, X_{i}   \}_{\alpha,\beta} \ &\equiv & \frac{1}{V_{\alpha\beta}} 
( \partial_{\alpha}
\dot{X}_{i}\partial_{\beta}\dot{X}_{i}- \partial_{\beta}\dot{X}_{i}\partial_{\alpha}X_{i}) \ = \ 0 , \nonumber \\ 
\ \alpha,\beta &=& 1,2,\ldots, p
\eeqa
where $V_{\alpha\beta}$ is the $\xi_{\alpha},\xi_{\beta}$ part of the volume element 
$d^{p}\xi \sqrt {\gamma}$ . 
%It has been noticed that the potential energy term of the Hamiltonian 
%can be rewritten in terms of the Nambu p-bracket 

%\beq
%det \ [\partial_{\alpha}X_{i} \partial_{\beta}X_{i} ] = \frac{1}{p!} \ \ \ \sum_{ %i_{1},\ldots, i_{p}=1}^{D-2}
%\{ X_{i_{1}},\ldots, X_{i_{p}}\}^{2}
%\eeq
 
%where
%\beq
%\{ f_{1},\ldots,f_{p}\} \equiv  \frac{1}{\sqrt{\gamma}} \epsilon^{\alpha_{1}\cdots %\alpha_{p}} \partial_
%{\alpha_{1}}f_{1}\cdots\partial_{\alpha_{p}}f_{p} \ \ \ \ \ \alpha_{1},\ldots, %\alpha_{p}=1,\ldots, p
%\eeq
The Nambu bracket is a generalization of the Poisson bracket of Classical Mechanics to "phase space" 
of any dimension p \cite{Nam}. It is a completely antisymmetric multilinear function of $f_{1}, \ldots, f_{p}$ and 
satisfies
two additional properties 
$(\alpha)$ Leibniz 
\beq
\{f_{1} \cdot g_{1}, f_{2}, \ldots , f_{p} \}= f_{1} \{ g_{1},f_{2}, \ldots , f_{p} \} + g_{1} 
\{ f_{1}, \ldots, f_{p} \}
\eeq
and 
$(\beta )$ the Fundamental Identity  which generalizes the Jacobi identity. 
Furthermore, it generalizes  
Lie algebras and Poisson Manifolds  to Nambu-Poisson and Nambu-Lie 
structures which turn out to be more rigid. 
\beqa
\{ \{f_{1}, f_{2},\ldots , f_{p} \}, f_{p+1}, \ldots, f_{2p-1} \}  & & \nonumber \\
 + \ \ \{ f_{p} , \{ f_{1} , f_{2} , \ldots, f_{p-1} , f_{p+1} \},
f_{p+2}, \ldots, f_{2p-1} \} + \ldots  \ & & \nonumber \\
 +  \ \ \{ f_{p},  , f_{p+1} , \ldots , 
f_{2p-2} , \{ f_{1} , f_{2} , \ldots , f_{p-1} , f_{2p-1} \} \}  &=&  \nonumber \\
\{ f_{1} , f_{2} , \ldots , f_{p-1} , \{ f_{p} , f_{p+1} , \ldots , f_{2p-1} \} \}  
\eeqa

There is one very interesting property 
of the Nambu bracket for spherical and toroidal p-branes . For $S^{p}$ - p-dim. branes of spherical topology there 
is a natural system
of functions $e_{1}, \ldots, e_{p+1}$ of the angles $ \Omega = (\phi, \theta_1, \theta_2, \ldots )$   
where a unit vector in the direction $\Omega $ in $p+1$ dimensional Euclidean space is expressed as

\beq
\hat{r} = ( e_{1}, \ldots,  e_{p+1} )
\eeq
 
with
 
\beq
 e_{1}^{2} +\ldots + e_{p+1}^{2} = 1 
\eeq
These functions (polar coordinates of   
$p+1$-vectors) can be easily checked to satisfy
\beq
\{ e_{i_{1}},\ldots, e_{i_{p}} \} = \epsilon_{i_{1}\cdots i_{p} i_{p+1} }\ e_{i_{p+1}} \ \ , \ \ \  \ \ 
i_{1}, \ldots , i_{p+1}= 1,\ldots , p+1
\eeq
For $p=2$ they are 

\beq
( e_{1},e_{2}, e_{3} ) = ( cos\phi sin\theta \ , \ sin\phi sin\theta \ ,\ cos\theta )
\eeq
Similarly for $p=3$ we have 
\beq
(e_{1}, e_{2}, e_{3}, e_{4} )= ( cos\phi sin\theta_{1} sin\theta_{2} \ ,\ sin\phi sin\theta_{1} sin\theta_{2}\ , \ 
cos\theta_{1} sin\theta_{2} cos\theta_{2} )
\eeq
The  corresponding volume elements are :

\beq
p = 2 \ \ \ \ \ \ \ d^{2}\Omega = sin \theta d \theta d \phi  \nonumber \\
\eeq
and similarly
\beq
p = 3 \ \ \ \ \   d^{3}\Omega = sin^{2}\theta_{2} sin\theta_{1} d\theta_{1} d\theta_{2} d\phi
\eeq
The Poisson and Nambu brackets are defined correspondingly as 
\beq
\{ f_{1} \ ,\ f_{2} \} \stackrel{p=2}{=} \frac{1}{sin\theta} ( \partial_{\theta} f_{1} \partial_{\phi} f_{2}
- \partial_{\phi} f_{1} \partial_{\theta} f_{2} )
\eeq 
  and
\beq
\{ f_{1}, f_{2} , f_{3} \} \stackrel{p=3}{=} \frac{1}{sin^{2}\theta_{2}sin\theta_{1}} \ \ \epsilon^{\alpha\beta\gamma} \ \
\partial_{\alpha}f_{1}\partial_{\beta}f_{2}\partial_{\gamma}f_{3}
\eeq
with  $ \alpha, \beta, \gamma = \theta_{1}, \theta_{2}, \phi $.
For the torus $T^{p}$ we have a flat measure for any p , $ d\omega= d\sigma_{1}\cdots d\sigma_{p} $ where
\beq
\sigma_{\alpha} \in ( 0, 2\pi ) , \ \ \ \ \alpha = 1,\ldots, p
\eeq
the basis functions are 

\beq
e_{\stackrel{\rightarrow}{n}} \ = \ e ^{i \stackrel{\rightarrow}{n} \cdot \stackrel{\rightarrow}{\sigma}} 
\ \ \ \ \ \stackrel{\rightarrow}{n} \in {\cal{Z}}^{p}
\eeq
while their Nambu brackets are 

\beq
\{ e_{\stackrel{\rightarrow}{n_{1}}},\ldots, e_{\stackrel{\rightarrow}{n_{p}}} \} \ = \ i^{p} \ \ det 
( \stackrel{\rightarrow}{n_{1}},\ldots,\stackrel{\rightarrow}{n_{p}}) \cdot 
e^{i( \stackrel{\rightarrow}
{n_{1}}+\cdots +\stackrel{\rightarrow}{n_{p}})\cdot \stackrel{\rightarrow}{\sigma}}
\eeq

Volume preserving transformation can be defined through the Nambu bracket. For fixed 
$f_{1},\ldots,f_{p-1}$ functions on the p-brane we define the generator

\beq
L_{(f_{1},\ldots,f_{p-1})} f \ = \ \{ f_{1},\ldots , f_{p-1}, f \}
\eeq
if f is functionally dependent on $f_{1},\ldots,f_{p-1}$ the result is zero. 
The operation is restricted to satisfy the 
fundamental identity (2.8).
As an example for the $3$-sphere $S^{3}$ for any two of the four functions 
$e_{1},e_{2},e_{3},e_{4}$  the operator

\beq
L_(e_{i},e_{j})f \ = \ \{ e_{i}, e_{j} , f \}
\eeq
executes a rotation on the plane i,j. In general if $ \alpha= \alpha_{i}\cdot e_{i},\  
 \beta= \beta_{j}\cdot e_{j}$ with  
($\alpha_{i},\beta_{j} \in \cal R $),

\beq
L_{\alpha ,\beta} \ \ f = \ \ \{ \alpha ,\beta, f \}
\eeq
executes a rotation in the plane $(\alpha,\beta)$.  In a future work we shall present the
structure of the algebras $(2.11)$ for $ S^{p}, T^{p}$. 

The case $p=2$ corresponds to the supermembrane and in this case there is a M(atrix) discretization
by Goldstone, Gardner, Hoppe \cite{Gold} which was revived in the late 80's \cite{Berg} and late 90's as the M(atrix) model \cite{Town}
proposal for M-theory. In the place of Poisson brackets one has commutators and in the place of target 
space $ X_{i}(\xi_{1}, \xi_{2}, t) , i=1,\ldots, D-2$ of membrane coordinates one has $N \times N$ Hermitian
matrices $ A_{i}(t)$ ( YM-mechanics in the Light Cone $10+1$ dimensions). In $3+1$ dimensions Yang-Mills mechanics 
was first studied by G.Savvidy \cite{Savv}. 
The equations of motion and constraints are given by :

\beq
\ddot{A_{i}} \ = \ - \left[  \ \ \left[ A_{i} , A_{j} \right], \  A_{i} \  \right] 
\ \ \ \ \ \ \ i,j=1, \ldots , D-2
\eeq
and
\beq
\left[ \dot{A_{i}} , A_{i} \right] \ = \ 0
\eeq
%We may observe that for both the p-branes and the M(atrix) Model the constraints generate the symmetries
%and hence commute with the hamiltonian. This means that given the initial conditions satisfying the 
%constraints the corresponding solutions of the eqs of motion satisfy the constraints for all times.
% The space-time symmetries of the LC eqs of motion are , galilean invariance generalized with rotations in 
% $ D-2 $ dimensions. 
 
% In this work we will be interested in  specific forms of  rotational motion, which correspond to lowest 
% energy configurations for p-branes and M(atrix) model. Bending or breathing modes cost more energy. 
% For this reason we call our solutions higher dimensional Euler Tops. 

For the case of factorization of the time ansatz\cite{AFP} it has been noticed that there is 
an isomorphism between the membrane $ p=2 $ and the matrix 
model solutions. As a consequence any $ p=2 $ 
spinning solution gives rise to a M(atrix) model solution.
In the next section we will find the
 conditions for this type of motion by generalizing the Euler eqs for Rigid Body Motion of classical 
 mechanics for p-branes in higher dimension.

%-----------------------------------------------------------------------------------------------------------
%where we have chosen as in \cite{peri1} the coordinates
%\beq
%Z_i \ = \ X_i + i X_{i+k}, \ \ i=1,..,k, \
%\eeq
%while we denote the rest of the space as follows :
%\beq
%Y_1 \ = \ X_7 , \ Y_2 \ = \ X_8 , \ Y_3 \ = \ X_9
%\eeq
%Then the equations of motion read :
%\beq
%\ddot{Z}_i = \{ \{Z_i, Z_j^{\star} \}, Z_j \} \ + \
%\{  \{ Z_i, Z_j \}, Z_j^{\star} \} \ + \
%\{ \{ Z_i, Y_a \}, Y_a \}
%\eeq
%The eqn's for the constraints read
%\beq
%\frac{1}{2}\{ \dot{Z}_i , Z_i^{\star}   \} \ + \
%\frac{1}{2}\{ \dot{Z}_i^{\star} , Z_i   \} \ + \
%\{ \dot{Y}_a^{\star} , Y_a  \}
%\eeq
%-----------------------------------------------------------------------------------------------------------
\newpage

\section{ P-Brane Euler Tops in Higher Dimensions }

In this chapter we derive the Euler eqs. for the purely rotational solutions of 
p-branes for any p. This type of motion presumably is the lowest in energy. 
Vibrational motion in radial or other directions costs more energy. Since p-branes
possess elastic tension their equilibrium shape is controlled , for purely rotational motion, 
by 
the balance between the rotational forces and tension. We will specify the necessary and
sufficient condition for
this equilibrium ansatz. 

The rotational or Euler Top motions of p-branes can be described by choosing some 
initial configuration $  X^{i}_{o}(\xi) $ with  $   
\xi= ( \xi_{1},\ldots, \xi_{p} )$ , and 

\beq
 X^{i}(t) \ = \  R^{ij} \ X_{o}^{j} ( \xi )\ \ , \ \ \ \ \ \ \ \ \  i,j =1, \ldots, D-2
\eeq 
where R is a time dependent rotation matrix,  $R \in SO(D-2)$ i.e. such that $ R^{T} = R^{-1} ,
R(t=0) \ = \ I $ the $ (D-2) \times (D-2) $ identity matrix. Let us introduce the
moments of inertial tensor in the brane frame 
\beq
 I_{B}^{ik} \ = \ T_{p} \int d^{p}\xi \ \sqrt{\gamma}  \ X^{i}_{o}(\xi ) \ \ X^{k}_{o}(\xi)\ \ , \ \ \ \ \ \
 i,k = 1,\ldots, D-2
 \eeq 
and the angular momentum tensor which is conserved in the fixed space coordinate frame 
 
 \beq
 L_{S}^{ij} \ = \ T_{p} \  \int d^{p}\xi  \ \sqrt{\gamma} \  \left(  \ \dot{X}^{i}X^{j} -
 \dot{X}^{j}X^{i} \right)\ \ , \ \ \ \ \ \ \ \ \ \ i,j=1, \ldots , D-2
 \eeq
%This is the fixed coordinate axes frame. 
%In order to connect the two frames we  introduce  $ I^{ij}_{S} $ given by
The two frames, Brane and Space, are connected through the Rotation Matrix R. 
We introduce the angular momentum in the Brane frame $ L_{B}$ and the Moment of 
Inertia in the Space Frame $ I_{S} $

\beq
 I_{S} \ = \ R(t) \cdot I_{B} \cdot R^{-1}(t) 
\eeq
and

\beq
 L_{B} \ = \ R^{-1}(t) \cdot L_{S} \cdot R(t) 
\eeq
The linking quantity between the angular momentum L and the moment of 
inertia tensor I is of course the angular velocity matrix in the two frames :

\beq
\omega_{S} \ = \ \dot{R} \ R^{-1} 
\eeq
and
 
\beq
\omega_{B} \ = \ R^{-1} \dot{R} 
\eeq
From the above definitions we obtain 
\beq
L_{B} \ = \ \omega_{B} \ I_{B} \ + \ I_{B} \ \omega_{B}
\eeq
and

\beq
L_{S} \ = \ \omega_{S} \ I_{S} \ + \ I_{S} \ \omega_{S} \ = \ R \cdot L_{B} 
\cdot R^{-1}
\eeq
From the conservation of $L_{S}$ we obtain 

\beq
\dot{L}_{B} \ + \ [ \omega_{B} , L_{B} ] \ = 0
\eeq
and from (3.8) the Euler eqs \cite{HGol}

\beq
\dot{\omega}_{B} I_{B} \ + \ I_{B} \dot{\omega}_{B} \ + \ 
[ \omega_{B}^{2} , I_{B} ] \ = \ 0 
\eeq
The above equation discloses the richness of rigid body dynamics generalized to 
higher dimensions \cite{HGol}. 
For the p-brane rotational motion we make the ansatz (3.1). 
The constraints impose the condition on $ \omega_{B}(t=0) \ = \ \omega_{B_{o}} $ 

\beq
\omega_{B_{o}}^{ij} \ \{ X^{i}_{o} , X^{j}_{o} \}_{\xi_{\alpha},\xi_{\beta}} \ = \ 0 
 \ \ , \ \ \ \ \ \ \ \ \  \alpha,\beta = 1, \ldots, p 
\eeq
This condition is easily satisfied if we partition the i,j range into a direct sum structure $(i_{1},j_{1}) , ( i_{2},j_{2} ), \ldots  $

\beq
\omega_{B_{o}} \ = \ \omega_{B_{o}}^{1} \oplus \omega_{B_{o}}^{2} \oplus  \cdots
\eeq
and impose $ \{ X_{o}^{i_{q}} , X_{o}^{j_{q}} \}_{\xi_{\alpha},\xi_{\beta}}
 \ = \ 0 $ , for all $q=1,2,\ldots  $ 
This is the general structure of our ansatz in the next chapters for $ S^{2} , S^{3} ,
T^{2}, T^{3} $ . 
On the other hand the eqs. of motion (2.5) produce the following additional constraints:

\beqa
 v^{ij}X^{j}_{o} \ &=& \ \frac{1}{(p-1) \ !} \{ \{ X^{i}_{o},X^{k_{1}}_{o}, \ldots, X^{k_{p-1}}_{o}\}, 
 X^{k_{1}}_{o},\ldots , X^{k_{p-1}}_{o} \} \\ & & 
 \ \ \ \ \ \ \ \ \ \ \ \ \ \ \ \ \ \ \ \ 
 \ \ \ \ \ \ i,k_{1},\ldots, k_{p-1} = 1,\ldots, D-2 \nonumber
 \eeqa
 where for all times 
 
 \beq
 v^{ij} \equiv ( R^{-1} \ddot{R} )^{ij}
 \eeq
and $X^{i}_{o}$ should close the algebra (3.14). In what follows, 
we are going to see that
this is guaranteed for special functions $X^{i}_{o}$.
The implication of rel.(3.15) is that

\beq
\ddot{R} \ = \ R \cdot v
\eeq
with $ R(t=0) = I \ , \ R^{T}R =I $  and $v$ is constant. The only solution to these
requirements is
\beq
R(t) \ = \ e^{\Omega \cdot t} \ \ \ , \ \ \  \Omega^{T}=-\Omega
\eeq
and thus $v$ is a symmetric non-negative definite matrix
\beq
v \ = \ \Omega^{2}
\eeq
The energetics of this ansatz goes as follows. The energy of the configuration
\beq
E\ =\ \frac{T_{p}}{2} \ \int\ d^{p}\xi \ \sqrt{\gamma} \ \left[ \ \dot{R}^{ij} \ X^{j}_{o}\ X^{k}_{o}\ \dot{R}^{ki} \ + \ 
\frac{1}{p \ !} \ \{ X^{i_{1}}_{o},\ldots, X^{i_{p}}_{o} \}^{2}\ \  \right] \
\eeq
consists of two conserved pieces:
The potential energy V 

\beq
V \ \ = \ \ \frac{T_{p}}{2p \ !} \ \int \ d^{p}\xi \ \sqrt{\gamma} \ \{ X^{i_{1}}_{o},\ldots, X^{i_{p}}_{o}\  \}^{2}
\eeq
and the kinetic energy which is expressed in terms of the conserved angular momentum 
( $L_{S}$ is minus the usual angular momentum) 
 \beq
 E_{kin}\ \ = \ -\frac{1}{2}\  tr \ \omega_{S}  \cdot  I_{S} \cdot \omega_{S}\  =\ -\frac{1}{4} 
 \ \ tr\ \ L_{S}\ \cdot \omega_{S}
 \eeq

 By integrating the equation of equilibrium of forces after multiplying by 
 $X^{i}_{o}$  eq. $(3.14)$  we get for the potential energy:
 \beq
 T_{p} \int d^{p}\xi \sqrt{\gamma} \ v^{ij}\ X^{i}_{o}\ \ X^{j}_{o} \ \ = \ \ -\frac{T_{p}}{(p-1)!}\ \ 
 \int d^{p}\xi \ \sqrt{\gamma} \  \{  X^{i_{1}}_{o}, \ldots, X^{i_{p}}_{o} \ \}^{2} \ \ = \ -2 p V 
 \eeq
 
 or 
 \beq
 tr \ v \cdot I_{B}\ \ =\ \ -2 p V 
 \eeq
 From (3.6-3.7) we obtain 
 
 \beq
 \omega_{B} \ = \ \omega_{S} \ = \Omega
 \eeq 
 and thus 
 \beq
 V \ = \ - \frac{1}{2p} \ \ tr \ \Omega^{2}I_{B}
 \eeq
 \beq
 E_{kin} \ = \ - \frac{1}{2} \ tr \ \Omega^{2} \ I_{B} \ = \ p V
 \eeq
 and
 \beq
 E_{tot} \ = \ - \left( \frac{1}{2} \ + \ \frac{1}{2p} \right) \ tr  \ 
 \Omega^{2} \  I_{B}
 \eeq
 Finally the relation of $E_{tot}$ to the conserved angular momenta is 
 \beq
 E_{tot} \ = \ - \ \frac{1}{4} \ \left( 1 \ + \ \frac{1}{p} \right) \ tr  \ \Omega L_{S}
 \eeq

\newpage

\section{Spherical P-Brane Tops ($S^{2}$ , $S^{3}$)}
\subsection{ $S^{2}$  Tops}

In this chapter we exhibit new spinning $p=2$ and $p=3$ spherical brane solutions
with rotational symmetries $ \prod_{i}SO(q_{i})$ . We render
transparent the role of the symmetry algebras which are formed by the Nambu-Poisson brackets and 
clarify the minimun energy character of the p-Euler Tops.
For  $S^{2}$ ($p=2$)  the relevant $SO(3)$ algebra for the basis functions 

\beqa
( e_{1},e_{2},e_{3}) \ &=& \ ( cos \phi sin \theta , sin\phi sin \theta , cos\theta ) 
 \nonumber  \\  \{ e_{i} , e_{j} \} \ &=& \ - \ \epsilon_{ijk}  \ e_{k}
\eeqa
is responsible for the polynomially generated universal enveloping algebra, the 
SDiff($S^{2}$).
It is known that the only finite dimensional subalgebras of SDiff($S^{2}$) is $SO(3)$.
Thus factorization with a finite number of time dependent modes can be found by using only the $e_{i}$s.
We propose a generalization of embeddings for $S^{2}$ in $9-dim$. 
\beq
R^{9}= R^{q_{1}} 
\times R^{q_{2}} \times R^{q_{3}} \ \ \ ,\ \ \ q_{1}+q_{2}+q_{3}=9 
\eeq
as follows :

\beqa
X_{i}  & \equiv &  x_{i}(t) \cdot e_{1}  \nonumber \\  
Y_{j} & \equiv &   X_{q_{1}+j} = y_{j} \cdot  e_{2} \nonumber \\ 
Z_{k} & \equiv &   X_{q_{1}+q_{2}+k} = z_{k} \cdot e_{3}   
\eeqa
 where ( $i,j,k = 1, \ldots, q_{1},q_{2},q_{3}$ ) \ respectively with
$q_{1}+q_{2}+q_{3}=9 $ and the $q_{i}$s are nonzero integers. The case 
$q_{1}=q_{2}=q_{3}=2$ for the matrix model has been studied in ref.\cite{Tayl,Harm} 
while for the membrane in ref\cite{AFP,Harm}. In principle 
one of the $q_{i}$, $i=1,2,3$ may be zero. The constraints 
\beq
\sum ^{9}_{i=1} \{ \dot{X}_{i} \ , \ X_{i} \} \ = \ 0
\eeq
are automatically satisfied. 

The functions $x_{i}, y_{j},z_{k}$  functions which determine the 
simultaneous time evolution of every point of $S^{2}$  in $ R^{9}$  satisfy the eqs. 
of motion
\beq
\ddot{\vec{x}} = - \vec{x} \ \  ( r_{y}^{2} \ +\ r_{z}^{2} \ )
\eeq

By cyclic  permutation on the $ x,y,z $  one obtains similarly
the eqs for $\vec{y}$ and $\vec{z}$
%\beqa
%\ddot{\vec{x}} &=& - \vec{x} \ \  ( r_{y}^{2} \ +\ r_{z}^{2} \ )\nonumber\\
%\ddot{\vec{y}} &=& - \vec{y} \ \  ( r_{x}^{2} \ + \ r_{z}^{2} \ ) \nonumber \\
%\ddot{\vec{z}} &=& - \vec{z} \ \  ( r_{x}^{2} \ + \ r_{y}^{2} \ ) 
%\eeqa
with $ \vec{x} = (x_{1},\ldots,x_{q_{1}})$ \ , \ $ \vec{y}=(y_{1},\ldots,y_{q_{2}})$ \ , \ $\vec{z}=
(z_{1}, \ldots, z_{q_{3}})$ and 
\beq
r_{x}^{2}= \sum_{i=1}^{q_{1}} x_{i}^{2} \ \ \  ,\ \ \ 
 r_{y}^{2} \ = \ \sum_{j=1}^{q_{2}} y_{j}^{2} \ \ \ , \ \ \ r_{z}^{2} \ = \ \sum_{k=1}^{q_{3}}
 z_{k}^{2}
\eeq
We see that eqs.$(4.3)$ admit an $SO(q_{1}) \times SO(q_{2}) \times SO(q_{3})\subset SO(9)$
rotational symmetry. 
The Hamiltonian of the ansatz 
\beq
H \ = \ \frac{T_{2}}{2} \ \ \int_{S^{2}} \ d^{2}\xi \ \ \left[ \dot{X}_{i}^{2} + \frac{1}{2}
\{ X_{i},X_j \}^{2} \right] 
\eeq
can be calculated by the use of the orthogonality relation 
\beq
\int_{S^{2}} d^{2}\xi \ \ e_{k} \ \cdot \ e_{l} \ = \ \frac{4 \pi}{3} \delta_{k,l} \ , \ \ \ k,l=1,2,3
\eeq
We find 
\beq
E\ =\ \frac{2\pi}{3} T_{2} \ \left[ \ \dot{\vec{x}}^{2} +\dot{\vec{y}}^{2}+\dot{\vec{z}}^{2}
+ r_{x}^2 r_{y}^{2}+r_{x}^2 r_{z}^{2}+r_{y}^2 r_{z}^{2} \right] 
\eeq
In order to relate the Energy with $SO(d_{1})$,$SO(d_{2})$,$SO(d_{3})$ 
angular momenta we observe that for each component separately we have
\beq
( L_{z})_{mn} \ = \ \frac{4\pi T_{2}}{3} \ ( \ l_{z} \ )_{m,n} \ \ \ , \ \ \ \ \  m,n = 1,\ldots, q_{3}
\eeq
 The same will hold true for $ (L_{y})_{kl}$ and $ (L_{x})_{ij}$ with 
 $ k,l = 1,\ldots, q_{2}$ and $ i,j = 1,\ldots, q_{1}$ respectively.

%\beq
%( L_{y})_{kl} \ = \ \frac{4\pi T_{2}}{3} \ ( \ l_{y} \ )_{k,l} \ \ \ \ 
%k,l = 1,\ldots, q_{2} 
%\eeq
%\beq
%( L_{x})_{ij}\ = \ \frac{4\pi T_{2}}{3} \ ( \ l_{x} \ )_{i,j} \ \ \  
%i,j = 1,\ldots, q_{1} 
%\eeq
Here $l_{x}$,$l_{y}$,$l_{z}$ are given by
\beq
( l_{x} )_{ij} \ = \ \dot{x}_{i} x_{j} \ - \ \dot{x}_{j} x_{i} 
\eeq
Similarly for $(l_{y})_{kl}$ and $(l_{z})_{mn}$.

The higher dimensional kinetic terms $ \dot{\vec{x}}^{2},\ldots $ can be expressed in 
terms of the radial and angular variables as :
\beq
\dot{\vec{x}}^{2} \ = \ {\dot{r}_{x}}^{2} \ + \ \frac{ l_{x}^{2}}{r_{x}^{2}} 
\eeq
Then the energy is given in terms of $l_{x}$,$l_{y}$,$l_{z}$ and $r_{x}$,$r_{y}$,$r_{z}$ as :

\beq
E = \frac{2\pi T_{2}}{3}  \left( E_{kin} \ + \ V_{eff} \right)
\eeq
where
\beqa
E_{kin}  &=&  \dot{r}_{x}^{2}  + \dot{r}_{y}^{2} + \dot{r}_{z}^{2} \nonumber \\
V_{eff}  &=&  \frac{l_{x}^{2}}{r_{x}^{2}}+\frac{l_{y}^{2}}{r_{y}^{2}}+
\frac{l_{z}^{2}}{r_{z}^{2}} + r_{x}^2 r_{y}^{2}+r_{x}^2 r_{z}^{2}+r_{y}^2 r_{z}^{2}
\eeqa
We are now in the position to make the connection between this ansatz and the 
Euler-Top formalism of ch.3. Due to the breaking of rotational symmetry $SO(9)$ to 
 $ SO(q_{1}) \times SO(q_{2}) \times SO(q_{3})$ the time-evolution of the vector 
 $ \vec{x}(t),\vec{y}(t),\vec{z}(t)$ is described by : 
 
 \beqa
 \vec{x}(t) \ &=& \ e ^{\Omega_{x} \cdot t} \ \vec{x}_{o} \nonumber \\
 \vec{y}(t) \ &=& \ e^{\Omega_{y} \cdot t} \ \vec{y}_{o} \nonumber \\
 \vec{z}(t) \ &=& \ e^{\Omega_{z} \cdot t} \ \vec{z}_{o}
 \eeqa
By using $ SO(q_{1})$, $ SO(q_{2})$, $ SO(q_{3})$ rotations we can bring the vectors
 $\vec{x}_{o}$, $\vec{y}_{o}$, $ \vec{z}_{o}$ to their corresponding first axes : 
 \beqa
  \vec{x}_{o}&=& R_{x} \ \ (1,\ldots, 0) \ \ \ q_{1} -\mbox{components} \ \ \ \nonumber \\
  \vec{y}_{o}&=& R_{y} \ \ (1,\ldots, 0) \ \ \ q_{2} -\mbox{components} \ \ \ \nonumber \\
  \vec{z}_{o}&=& R_{z} \ \ (1,\ldots, 0) \ \ \ q_{3} -\mbox{components}
 \eeqa
By keeping the position vectors fixed we can bring the initial velocities to the planes
 $ x^{1} x^{2} $ , $ y^{1}y^{2} $ , $ z^{1}z^{2} $.  
 Thus,  each  $ \Omega_{i} \ \ ( i=x,y,x) $    angular velocity matrix becomes
 
 \beq
 \Omega_{i} \ = \ \left( \ba{cc} 0 & -\omega_{i} \\ \omega_{i} & 0 \ea \right)
 \eeq
 %\beqa
 %\Omega_{x} \ = \ \left( \ba{cc} 0 & -\omega_{x} \\ \omega_{x} & 0 \ea \right) \nonumber \\
 %\Omega_{y} \ = \ \left( \ba{cc} 0 & -\omega_{y} \\ \omega_{y} & 0 \ea \right) \\
 %\Omega_{z} \ = \ \left( \ba{cc} 0 & -\omega_{z} \\ \omega_{z} & 0 \ea \right) \nonumber
 %\eeqa
in their respective planes  and zero for all others. 
The moment of inertia tensor acquires a similar
 form: 
 
 \beqa
 I_{B} &=& I_{x} \oplus I_{y} \oplus I_{z} \nonumber \\
 I_{i} &=& \frac{2\pi T_{2}}{3} \ \ \left( \ba{cc} R_{i}^{2} & 0 \\ 0 & 0 \ea \right) \ ,\ \ \ \ 
 \ \ \ i=x,y,z
 \eeqa
and so $ L_{B} = L_{S} $  where 
 
 \beqa
 L_{B} &=& \omega_{x} I_{x} \left( \ba{cc} 0 & -1 \\ 1 & 0 \ea \right) \oplus 
 \omega_{x} I_{x} \left( \ba{cc} 0 & -1 \\ 1 & 0 \ea \right)\oplus 
 \omega_{x} I_{x} \left( \ba{cc} 0 & -1 \\ 1 & 0 \ea \right) \nonumber \\
  & \equiv & L_{x} \oplus L_{y} \oplus L_{z}
 \eeqa
The total energy according to rel.$(3.28$) is : 
 
 \beq
 E \ = \ \frac{1}{2} \left( \omega_{x}^{2} I_{x} + \omega_{y}^{2} I_{y}+ \omega_{z}^{2} I_{z} \right)
 \eeq
The balance of force condition  relates the angular momenta with the 
radii of rotation
as
\beq
  \omega_{x}^{2} = R_{y}^{2} + R_{z}^{2}
\eeq
Similarly for $\omega_{y}$ and $\omega_{z}$.
 
 %\beqa
 %\omega_{x}^{2} &=& R_{y}^{2} + R_{z}^{2} \nonumber \\
 %\omega_{y}^{2} &=& R_{x}^{2} + R_{z}^{2} \\
 %\omega_{z}^{2} &=& R_{x}^{2} + R_{y}^{2} \nonumber
 %\eeqa
These equations are identical to the ones obtained from the minimization of the effective
  potential $V_{eff}$ ($4.14$) which lead to constant radii solutions :
  
  \beq
  r_{i} = R_{i}\ \ , \ \ \ \ \ \ \ \ \ \ i =x,y,z
  \eeq
We now proceed to present details of the solutions of the minimization conditions which
 provide an interesting complex dependence of the Energy $(4.20)$  on the angular momenta 
 $L_{x},L_{y}.L_{z}$. The extrema of the Energy are given by constant in time radii $r_{x},r_{y},
 r_{z}$ satisfying :
 
 \beq
 \frac{\partial V_{eff}}{\partial r_{x}} \ = \ -\frac{2 l_{x}^{2}}{r_{x}^{3}} \ + \ 2 r_{x} (r_{y}^{2}
 + r_{z}^{2}) \ = \ 0
 \eeq
 and so on for $r_{y},r_{z}$. The system of equations to be solved are :
 \beq
 r_{x}^{4} \ ( \ r_{y}^{2} \ + \ r_{z}^{2} )\ = \ l_{x}^{2}
 \eeq
 The rest can be obtained by permutation symmetry 
$ x \leftrightarrow y, l_{x} \leftrightarrow l_{y},\ldots \mbox{etc}$ 
 
%\beqa
 %r_{x}^{4} \ ( \ r_{y}^{2} \ + \ r_{z}^{2} )\ &=& \ l_{x}^{2} \nonumber \\
 %r_{x}^{4} \ ( \ r_{y}^{2} \ + \ r_{z}^{2} )\ &=& \ l_{x}^{2} \\
 %r_{x}^{4} \ ( \ r_{y}^{2} \ + \ r_{z}^{2} )\ &=& \ l_{x}^{2} \nonumber
 %\eeqa
 which can be solved for general $l_{x}^{2}, l_{y}^{2}, l_{z}^{2}$ . 

%By denoting $s_{1,2,3}$ :
 
 %\beqa
 %s_{1} \  &=&  \ r_{x}^{2} \  +  \ r_{y}^{2} \  + \  r_{z}^{2} \nonumber \\
 %s_{2} \ &=&  \ r_{x}^{2} \ r_{y}^{2} \  + \  r_{x}^{2} \ r_{z}^{2} \\
 %s_{3} \ &=&  \ r_{x}^{2} \ r_{y}^{2} \ r_{z}^{2} \nonumber
 %\eeqa
%and by $ \alpha_{1},\alpha_{2}, \alpha_{3}$ : 
 
% \beqa
% \alpha_{1} \ &=& \ l_{x}^{2} \ + \ l_{y}^{2} \ + \ l_{z}^{2} \nonumber \\
% \alpha_{2} \ &=& \ l_{x}^{2} \ l_{y}^{2} \ + \ l_{x}^{2} \ l_{z}^{2}+ l_{y}^{2} \ l_{z}^{2} \\
% \alpha_{3} \ &=& \ l_{x}^{2} \ \ l_{y}^{2} \ \ l_{z}^{2} \nonumber
% \eeqa
%we find that $s_{1},s_{2},s_{3}$ satisfy the eqs.:
 
% \beqa
% 2 s_{3}^{3} + \alpha_{1} s_{3}^{2} &=&  \alpha_{3}  \nonumber \\
% s_{1} s_{2} - 3 s_{3} &=& \alpha_{1}\\
% s_{2}^{3} + 3 s_{3}^{2} - 3 s_{1}s_{2}s_{3} &=& \alpha_{2} \nonumber
% \eeqa
%and the minimun of the potential is given as :
 
% \beq
% V_{eff}^{min} \ = \ 3 \cdot s_{2} \ = 3\ \left( \alpha_{2} \ + \ 2 \alpha_{1} s_{3} \ + \ 
% 3 s_{3}^{2} \right)^{1/3}
% \eeq
%Although $s_{3}$ satisfies a cubic equation $(4.2)$ so that all of $s_{1},s_{2},s_{3}$ can be determined
% and $r_{x}^{2},r_{y}^{2},r_{z}^{2}$ are the real positive roots of 
% \beq
% x^{3} \ - \ s_{1} \ x^{2} \ + \ s_{2} \ x \ -s_{3} \ = \ 0
% \eeq
We exhibit solutions only for the two simplest cases :
 \beq
 l_{i}^{2} \ \equiv l^{2} \ \ \ , \ \ \ r_{i}^{2} \ \equiv \ r^{2} \ \ \ , \ \ \ i=x,y,z
 \eeq
the completely symmetric case (S) and 
 
\beq
l_{x}^{2} \ = \ l_{y}^{2} \ = \ l^{2}  \neq  l_{z}^{2} \ \ \ , \ \ \ r_{x}^{2} \ = \ r_{y}^{2} \ = \ 
 r_{\alpha}^{2} \ \neq \ r_{z}^{2} 
\eeq
the axially symmetric case (A).
 Before that though we will demonstrate that the extrema $(4.22)$ are local minima of the energy. 
 Indeed, by taking the second variation of the potential at the extrema \cite{AFP,Harm};
 \beq
 \frac{ \partial^{2} V}{\partial r_{i} \partial r_{j}} \ \bigg|_{i,j=x,y,z} \ = \ 
 4 \ \ \left( \ba{ccc} 2(r_{y}^{2}+r_{z}^{2}) &  r_{x}r_{y} & r_{x}r_{z} \\
 r_{y}r_{x} &  2(r_{x}^{2}+r_{z}^{2}) &  r_{y}r_{z} \\ r_{z}r_{x} & r_{z}r_{y}& 
 2(r_{x}^{2}+r_{y}^{2})
 \ea \right)
 \eeq
we check that this is a real symmetric matrix (real eigenvalues) but also positive definite 
 (positive eigenvalues) i.e. for arbitrary real vectors  \ \ $\xi_{i} \in  \cal{R}$  , $ i=x,y,z$ we
 find 
 
 \beq
 \xi_{i} \ \xi_{j} \ \frac{\partial^{2} V}{\partial r_{i} \partial r_{j}} \ \ > \ 0 
 \eeq
We shall compare now, energy wise, the symmetric with the axially symmetric case $(4.25)-(4.26)$. 
 For the symmetric case (S) we find 
 
 \beq
 r_{S}^{2} \ \equiv \ r^{2} \ = \ \left( \frac{l^{2}}{2} \right) ^{1/3} \ \ \ , \ \ 
 V_{eff}^{min}\ = \ V_{S} \ =
 \ \frac{9}{4^{1/3}} \cdot (l^{2})^{2/3} 
 \eeq
For the axially symmetric case we find 
 
 \beqa
 r_{z}^{2} \ &=& \ \frac{l_{z}}{2 l^{2/3}} \ \left( l_{z} \ + \ \sqrt{ l_{z}^{2}+8 l^{2}}\right)^{1/3} \nonumber \\
 r_{\alpha}^{2} \ &=& \ \frac{2 l^{4/3}} {\left( l_{z} \ + \ \sqrt{l_{z}^{2}+8 l^{2}}\right)^{2/3}}  \\
 V_{eff}^{min} &\equiv &  V_{\alpha} = \frac{6 l^{2/3}}{\left( l_{z} + \sqrt{l_{z}^{2} + 8 l^{2}}\right)^{4/3}}
   \ \ \left[ l_{z} ( l_{z}+ \sqrt{l_{z}^{2}+ 8 l^{2}}) + 2 l^{2} \right]  \nonumber
 \eeqa
In order to compare the two minima we rescale $ l_{z}=\lambda \ l $ and we identify it (l) 
in each of the two cases . We find for the ratio

\beq
\frac{ V_{\alpha}}{V_{s}} \ = \ f(\lambda ) \ = \ \frac{2^{5/3}}{3} \ \ \
 \frac{\lambda \left( \lambda + \sqrt{\lambda^{2}+8}\right) + 2 }{\left( \lambda + \sqrt{\lambda^{2}+8}\right)^{4/3}}
 \eeq
while for the radii : $ r_{s} = \left( \frac{l^{2}}{2} \right) ^{1/3}$ 

\beqa 
\frac{r_{z}^{2}}{r_{s}^{2}} \ \ &=& \ \ \frac{\lambda}{2^{2/3}} \ \left( \lambda + \sqrt{\lambda^{2}
+ 8} \right)^{1/3} \nonumber \\ 
 \frac{r_{\alpha}^{2}}{r_{s}^{2}} \ \ &=& \ \frac{2^{4/3}}{\left( \lambda + \sqrt
{\lambda^{2}+8} \right)^{2/3}}
\eeqa
where for 
$\lambda \rightarrow 1$ , $ f(\lambda) \rightarrow 1$ , and 
$ r_{z}^{2}/r_{s}^{2} \ = \ r_{\alpha}^{2} / r_{s}^{2} \ = \ 1 $ 

%As for the two other limits $ \lambda \rightarrow \infty , \lambda \rightarrow  0 $ 
%we obtain 

%\beqa 
%f(\lambda ) \ \ & \stackrel{\lambda\rightarrow \infty}{\longrightarrow} & \ \ \
%\frac{2^{5/3}}{3} \ \lambda^{2/3} \nonumber \\
%\frac{r_{z}^{2}}{r_{s}^{2}} \ \ & \stackrel{\lambda\rightarrow \infty}{\longrightarrow}& 
%\frac{1}{2^{2/3}} 
%\lambda^{2/3} \nonumber \\
%\frac{r_{\alpha}^{2}}{r_{s}^{2}} \ \ &\stackrel{\lambda\rightarrow \infty} {\longrightarrow} &
% 2^{4/3} \lambda ^{-2/3} 
%\eeqa
%and for  $( \lambda \rightarrow 0 )$ we  obtain 
 
%\beqa 
%f(\lambda ) \ \ & \stackrel{\lambda\rightarrow 0}{\longrightarrow} & \ \ \
%\frac{1}{3 \cdot 2^{1/3}} \ < \ 1 \nonumber \\
%\frac{r_{z}^{2}}{r_{s}^{2}} \ \ & \stackrel{\lambda\rightarrow 0 }{\longrightarrow}& 
%2^{1/3} \ \lambda  \nonumber \\
%\frac{r_{\alpha}^{2}}{r_{s}^{2}} \ \ & \stackrel{\lambda\rightarrow 0 }{\longrightarrow} &  
%\frac{1} {2^{2}}
%\eeqa
From the above analysis we deduce that if the membrane length in one dimensionality 
( say $q_{3}$ ) is much bigger than the 
 other two $(q_{1},q_{2}) $ it looses energy with respect to the symmetric case, while if 
 it is much smaller than the other two it gains energy. 
 The expression of the Energy as a function of the angular momenta and tension shows the 
 non-perturbative character of the spinning solutions. It also affords us the possibility
to quantize the rotational modes of the $S^{2}$ membrane by using $L^{2}$ and $L_{z}^{2}$ 
as Casimirs ( with eigenvalues $ \hbar n(n+q-2),\ \ n=0,1,2,\ldots $ for $SO(q)$ of the 
 $SO(q_{1}), SO(q_{2}), SO(q_{3})$ rotational groups ). 
The classical $S^{2}$ spinning 
 membranes live in a 6-dims out of the total nine while the quantum one occupies all 
 dimensions due to the rotational wave functions $SO(q_{1}), SO(q_{2}), SO(q_{3})$ (spherical harmonics)
in $q_{1}+q_{2}+q_{3}=9 $ dimensions. Concerning the stability of our solution,
 as we have already shown, there is classical and quantum mechanical perturbative stability
for the radial modes and quadratic expansion in $r_{x},r_{y},r_{z}$ around the minima will
exhibit the perturbative vibrational spectrum. Stability for the multipole in $ \theta, \phi$ 
fluctuations exists only for the symmetric case $ l_{x}=l_{y}=l_{z}$ as can be shown 
 by using the results of \cite{AFP,Harm}. 
The geometry of the ansatz with rotating axes is that of an ellipsoid 
which at any time satisfies the eqs:

\beq
\frac{ \vec{X}^{2}}{r_{x}^{2}} \ \ \ + \ \ \ \frac{\vec{Y}^{2}}{r_{y}^{2}} \ \ \ + \ \ \ 
\frac{\vec{Z}^{2}}{r_{z}^{2}} \ \ =\ \ 1
\eeq
On the other hand by suitable rotations only three planes survive ,i.e. $(12)$ of 
 the $q_{1}$, 
 $q_{2}$ and  $q_{3}$ dimensions respectively. Thus the two dimensional $S^{2}$ surface is 
 moving in a fixed 5-dimensional ellipsoid in the 9-dim space. We can use this observation to 
 argue for M-theory curved gravitational backgrounds with 5-dimensional ellipsoidal minimal 
 submanifolds (pp waves for example). Our spinning solutions
are isometrically embedable in these backgrounds ,i.e. they satisfy eqs. of motion in these 
 backgrounds.

\subsection{ $S^{3}$  \ \ Tops}

We close this section by presenting new spinning  $S^{3}$-brane  solutions. The branes for $p=3$ 
attract a lot of attention due to their possible role as fundamental particles , 
 YM-Gravity dualities \cite{Mal,Mets}, Matrix Cosmology \cite{FGS,Gibb}, giant gravitons
\cite{Mal,GSToum} etc. 
Although pp-waves with fluxes present interesting 
backgrounds, we will hereby consider only flat LC-spacetimes in order to show that
local minima of the energy can be found by appropriately balancing , generalizing spinning solutions, 
rotation with tension forces in this case too. The Hamiltonian
for an $S^{3}$ brane (see ch.2) in LC gauge can be written in terms of the Nambu 3-brackets

\beq
H \ = \ \frac{T_{3}}{2} \ \ \int d\Omega_{3} \  \left[ \ \dot{X}^{{i}^{2}} \ + \ \frac{1}{3!} \ \{ X^{i},X^{j},
X^{k} \}^{2} \right]
\eeq       
so that the resulting equations of motion and constraints are : 
 
 \beqa
 \ddot{X}^{i} &=& \frac{1}{2}  \{ \{ X^{i}, X^{j}, X^{k} \}, X^{j}, X^{k} \} \ \ ,\ \ \ \ \ \ \ \ 
 i,j,k = 1 , \ldots , d \leq D-2 \nonumber \\
  \{   \dot{X}^{i}  ,  & X^{i}& \}_{ \xi_{\alpha}, \xi_{\beta} } \ = \ 0 \ \ \ , \ \ \ \ \ \ 
 \ \ \ \ \ \ \ \ \ \ \ \ \ \ \ \ \ \ \alpha \neq \beta = \theta, \phi,\psi
 \eeqa 
with 
 
 \beqa
 d\Omega_{3} \ &=& \ sin^{2}\psi \ sin\theta \ d\psi \ d\theta \ d\phi \nonumber \\
 (\xi_{\alpha}) &=& ( \theta , \phi , \psi ) \ , \ \ \ \ 0 \leq \theta, \psi \leq \pi , 
 0 \ \leq \ \phi \ \leq \ 2 \pi
 \eeqa
and the Nambu 3-bracket for $S^{3}$ : 
 
 \beq
 \{ X^{i},X^{j},X^{k} \} \ = \ \frac{-1}{sin^{2}\psi \ sin\theta} \ \cdot \ \epsilon^{\alpha\beta\gamma}
 \partial_{\alpha}X^{i}\partial_{\beta}X^{j}\partial_{\gamma}X^{k} , \ \ \ \ \ \ \xi_{1}=\theta , 
  \xi_{2}=\phi , \xi_{3}=\psi 
 \eeq
As we discussed in ch.2 for $ S^{3}$ ( here $p=3$ ) there are $ p+1=4$ functions 
 $( e_{1}^{2}+ e_{2}^{2} + e_{3}^{2} + e_{4}^{2} = 1)$ 
 
 \beqa
 e_{1} \ &=& \ cos \phi \ sin \theta \ sin \psi \nonumber \\
 e_{2} \ &=& \ sin \phi \ sin \theta sin \psi \nonumber \\
 e_{3} \ &=& \ cos \theta sin \psi \nonumber \\
 e_{4} \ &=& \ cos \psi 
 \eeqa
closing the Nambu-bracketed (volume preserving $S^{3}$, Diff's) algebra, here global $SO(4)$ 
 rotations 
 
 \beq
 \{ e_{\alpha} , e_{\beta}, e_{c} \} \ = \ - \epsilon_{\alpha\beta c d} e_{d} \ ,\ \ \ \ \ \ 
 \alpha , \beta ,c,d = 1,2,3,4
\eeq 
As is the case with $S^{2}$ this is crucial for the factorization of time and $ \theta,\phi,\psi$ 
 dependence of the eqs. of motion. Thus with an analogous to $S^{2}$ ansatz  satisfying the constraints
 $(4.43)$ 
 
 \beqa
 X^{i} \ \ &=& \ \ x^{i}(t) \ e_{1} \ \ , \ \ \ \  i=1,\ldots, q_{1} \nonumber \\
 Y^{j} \ \ &=& \ \ X^{j+q_{1}} \ \ = \ \ y^{j(t)}e_{2} \ , \ \ \ j=1,\ldots, q_{2} \nonumber \\
 Z^{k} \ \ &=& \ \ X^{k+q_{1}+d_{2}} \ \ = \ \ z^{k}(t) e_{3} \ , \ \ \ k=1,\ldots, q_{3} \nonumber \\
 W^{l} \ \ &=& \ \ X^{l+q_{1}+q_{2}+q_{3}} \ \ = \ \ w^{l}(t) e_{4} \ , \ \ \ \ l=1 ,\ldots, q_{4} 
 \eeqa
with
 $q_{1}+q_{2}+q_{3}+q_{4} = d \leq D-2 $ , $ q_{\alpha} \geq 0 $  , $ \alpha = 1,2,3,4 $ 
we obtain: 
  \beq
  \ddot{\vec{x}} \ = \ - \vec{x} \ ( r_{y}^{2}r_{z}^{2} + r_{y}^{2} r_{w}^{2} + 
   r_{z}^{2}r_{w}^{2} )
  \eeq
 By cyclic permutation one obtains similarly the eqs of motion for 
 $ \vec{y} ,\vec{z}, \vec{w} $. 
 %\beqa
 %\ddot{\vec{x}} \ &=& \ - \vec{x} \ ( r_{y}^{2}r_{z}^{2} + r_{y}^{2} r_{w}^{2} + 
 %r_{z}^{2}r_{w}^{2} ) 
 %\nonumber \\ 
 %\ddot{\vec{y}} \ &=& \ - \vec{y} \ ( r_{x}^{2}r_{z}^{2} + r_{x}^{2} r_{w}^{2} + 
 %r_{z}^{2}r_{w}^{2} )\nonumber \\ 
 %\ddot{\vec{z}} \ &=& \ - \vec{z} \ ( r_{x}^{2}r_{y}^{2} + r_{x}^{2} r_{w}^{2} + 
 %r_{y}^{2}r_{w}^{2} )\nonumber \\
 %\ddot{\vec{w}} \ &=& \ - \vec{w} \ ( r_{x}^{2}r_{y}^{2} + r_{x}^{2} r_{z}^{2} + 
 %r_{z}^{2}r_{y}^{2} ) 
 %\eeqa
where $r_{x},r_{y},r_{z},r_{w}$ are the lengths of the vectors $\vec{x},\vec{y},\vec{z},
 \vec{w}$ respectively. From $4.40$ we see that the rotational symmetry $SO(d)$ is 
 broken down to $ SO(q_{1}) \times SO(q_{2}) \times SO(q_{3}) \times SO(q_{4})$. Of course 
 we must have $q_{i} \geq 2 , i=1,2,3,4$ in order to have at least $SO(2)$ rotational symmetry. 
 Otherwise (i.e. if some $q_{\alpha}=1$) we have less rotational symmetry. 
 The Energy-Angular momenta of the ansatz are: 

 \beqa    
 E &=& \frac{T_{3}} {2}  \frac{\mbox{Vol}(S^{3})}{4} [ \dot{r}_{x}^{2} \ + \ 
 \dot{r}_{y}^{2} \ +\dot{r}_{z}^{2} \ + \dot{r}_{w}^{2} \ + 
 \frac{l_{x}^{2}}{r_{x}^{2}} \ + 
 \frac{l_{y}^{2}}{r_{y}^{2}} \ +\frac{l_{z}^{2}}{r_{z}^{2}} \ + 
 \frac{l_{w}^{2}}{r_{w}^{2}} 
 \nonumber \\ & + &  r_{x}^{2} \ r_{y}^{2} \ r_{z}^{2} \ + \   
  r_{x}^{2} \ r_{y}^{2} \ r_{w}^{2} \ + \  
  \ r_{y}^{2} \ r_{z}^{2} \ r_{w}^{2} \ + \ 
 r_{x}^{2} \ r_{z}^{2} \ r_{w}^{2} \  ]
 \eeqa
 where $ \mbox{Vol}(S^{3}) = 2 \pi^{2} $ and the angular momenta are 
 
 \beqa
 L_{i}^{2} \ &=& \ \left( \frac{T_{3} \pi^{2}}{2 \cdot 2} \right)^{2} \ \ l_{i}^{2} 
 \ , \ \ \ \ i=x,y,z,w 
%\nonumber \\
% L_{y}^{2} \ &=& \ \left( \frac{T_{3} \pi^{2}}{2 \cdot 2} \right)^{2} \ \ l_{y}^{2} %\nonumber \\
% L_{z}^{2} \ &=& \ \left( \frac{T_{3} \pi^{2}}{2 \cdot 2} \right)^{2} \ \ l_{z}^{2} %\nonumber \\
% L_{w}^{2} \ &=& \ \left( \frac{T_{3} \pi^{2}}{2 \cdot 2} \right)^{2} \ \ l_{w}^{2} 
 \eeqa
and 
  \beq
  l_{x}^{2} \ \ = \ \ \sum_{i \neq j = 1}^{q_{1}} \ \ ( \ \dot{x_{i}} x_{j} \ - 
  \ \dot{x_{j}} x_{i})^{2}
  \eeq
 Similarly for $ l_{y}, l_{y}, l_{z}$

 %\beqa
 %l_{x}^{2} \ \ &=& \ \ \sum_{i \neq j = 1}^{q_{1}} \ \ ( \ \dot{x_{i}} x_{j} \ - 
 %\ \dot{x_{j}} x_{i})^{2} \nonumber \\
 %l_{y}^{2} \ \ &=& \ \ \sum_{i \neq j = 1}^{q_{2}} \ \ ( \ \dot{y_{i}} y_{j} \ - 
 %\ \dot{y_{j}} y_{i})^{2} \nonumber \\
 %l_{z}^{2} \ \ &=& \ \ \sum_{i \neq j = 1}^{q_{3}} \ \ ( \ \dot{z_{i}} z_{j} \ - 
 %\ \dot{z_{j}} z_{i})^{2} \nonumber \\
 %l_{w}^{2} \ \ &=& \ \ \sum_{i \neq j = 1}^{q_{4}} \ \ ( \ \dot{w_{i}} w_{j} \ - 
 %\ \dot{w_{j}} w_{i})^{2} \nonumber \\
 %\eeqa
If all $ l_{x,y,z,w}^{2}$ are different from zero the minimization condition for the 
 $V_{eff}$ is equivalent to constant radii solutions 
 \beq
 l_{x}^{2} \ = \ r_{x}^{4} \ ( \ r_{y}^{2} r_{z}^{2} + r_{y}^{2} r_{w}^{2} + 
 r_{z}^{2} r_{w}^{2})
 \eeq

 %\beqa
 %l_{x}^{2} \ &=& \ r_{x}^{4} \ ( \ r_{y}^{2} r_{z}^{2} + r_{y}^{2} r_{w}^{2} + 
 %r_{z}^{2} r_{w}^{2}) \nonumber \\
 %l_{y}^{2} \ &=& \ r_{y}^{4} \ ( \ r_{x}^{2} r_{z}^{2} + r_{x}^{2} r_{w}^{2} + 
 %r_{z}^{2} r_{w}^{2}) \nonumber \\
 %l_{z}^{2} \ &=& \ r_{z}^{4} \ ( \ r_{x}^{2} r_{y}^{2} + r_{x}^{2} r_{w}^{2} + 
 %r_{y}^{2} r_{w}^{2}) \nonumber \\
 %l_{w}^{2} \ &=& \ r_{w}^{4} \ ( \ r_{x}^{2} r_{y}^{2} + r_{x}^{2} r_{z}^{2} + 
 %r_{z}^{2} r_{y}^{2}) 
 %\eeqa
Indeed, we can check that these are local minima 
( $ \frac{\partial^{2} V}{\partial r_{\alpha}
 \partial r_{\beta}} \arrowvert_{minima} $   
 is positive definite). 
 With analogous arguments with the $S^{2}$ case the minimization condition can be solved due 
 to permutation symmetry 
 $( x \rightarrow  y \rightarrow  z \rightarrow  w )$ with,in general, fourth order polynomial
 equations.We will exhibit, in what follows,  the two simplest cases : (a) symmetric , $ r_{x}=r_{y}=r_{z}=r_{w}
 =R $, $ l_{x}=l_{y}=l_{z} = l_{w} = l $, and  (b) axially symmetric $ r_{x}=r_{y}=r_{z}=R $,
 $ r_{w}=R_{w}$, and $ l_{x}=l_{y}=l_{z}=l , l_{w}$ .
 For the symmetric case we get :
 
\beq
 R_{sym}^{2} \ = \  \left( \frac{ l^{2} }{3} \right) ^{1/4} 
 \eeq
 
 \beq
 E_{sym} \ = \ 2 T_{3} Vol(S^{3}) \ \left( \frac{l^{2}}{3}\right)^{3/4}
 \eeq
 For the axisymmetric case the radii are: 
 
 \beqa
 R^{2} \ &=& \ \left( \frac{l_{w}^{2}}{3} \right)^{1/4} \ \ \left[ \sqrt{1 \ + \ 3 \frac{l^{2}}
 {l_{w}^{2}}} -1\right]^{1/2} \nonumber \\ 
 R_{w}^{2} \ &=& \ \frac{ ( l_{w}^{2} /3 )^{1/4} }{  [ \sqrt{1 + 3  l^{2}/l_{w}^{2}} - 1 ]^{1/2}}
 \eeqa 
 and the energy 
 
 \beq
 E_{ax}\ = \ \frac{T_{3}\mbox{Vol}(S^{3})}{2} \left(\frac{l_{w}^{2}}{3} \right)^{3/4} \ \  \left[ 2 \ + \ 
 \sqrt{ 1 \ + \ 3 \frac{l^{2}}{l_{w}^{2}}} \right] \ \ \left[ \sqrt{ 1 \ + \ 
 3 \frac{l^{2}}{l_{w}^{2}}} - 1 \right]^{1/2}  
 \eeq
 By rescaling $ l_{w}^{2} = \lambda l^{2} $ we find 
 
 \beq
 \frac{E_{ax}}{E_{sym}} \ = \ \frac{\lambda^{3/4}}{4} \ \ \left( 2 \ + \ \sqrt{1 + \frac{3}{\lambda}} \right)  
 \left( -1 \ + \ \sqrt{1 + \frac{3}{\lambda}} \right)^{1/2}
 \eeq
 also
 \beq
 \frac{R_{w}^{2}}{R^{2}} \ = \ \frac{1}{\sqrt{ 1 \ + \ \frac{3}{\lambda}} - 1} 
 \eeq
 For $\lambda = 1$ we have the symmetric case. For $\lambda \rightarrow 0 $ , we find qualitatively similar results with $S^{2}$ : 
 For $\lambda \rightarrow \infty $ we find 
 
 \beqa
 \frac{E_{ax}}{E_{s}} \ \ &\stackrel{\lambda\rightarrow 0}{\longrightarrow}& \ \ \ \frac{3^{3/4}}{4}
 \ \ \ <  \ \ \ 1  \nonumber \\
 \frac{E_{ax}}{E_{s}} \ \ & \stackrel{\lambda\rightarrow \infty}{\longrightarrow}&  \ \ \ \lambda^{1/4}
  \ \ \ \frac{3}{2^{2}}(\frac{3}{2})^{1/2} \ \ > \ \ 1
  \eeqa
 As far as the time dependence is concerned we can choose without loss of generality 4-planes
  $x^{1}x^{2}, y^{1}y^{2}, z^{1}z^{2}, w^{1}w^{2} $ where the initial position and velocity vectors
  belong. Then the ansatz $(4.47)$ of constant radii (at the minima) $ r_{x}=R_{x}, r_{y}=R_{y}, 
  r_{z}=R_{z}, r_{w}=R_{z} $ 
  
  \beq
  \dot{\vec{x}}(t) \ = e^{\Omega_{x} t} \vec{x}(0)
  \eeq
  Similarly for  $\dot{\vec{y}}, \dot{\vec{z}}, \dot{\vec{w}}$.

  %\beqa
  %\dot{\vec{x}}(t) \ &=& e^{\Omega_{x} t} \vec{x}(0) \nonumber \\
  %\dot{\vec{y}}(t) \  &=& e^{\Omega_{y} t} \vec{y}(0) \nonumber \\
  %\dot{\vec{z}}(t) \  &=& e^{\Omega_{z} t} \vec{z}(0) \nonumber \\
  %\dot{\vec{w}}(t) \  &=& e^{\Omega_{w} t} \vec{w}(0) 
  %\eeqa
  with \ \ \ $ \Omega_{i} \ = \ \left( \ba{cc} 0 & -w_{i} \\ w_{i} & 0 \ea \right)$
  ,\ \ \  $ i \ = \ x,y,z,w $ \ \ \ and the balancing of force conditions give \ \ \ \ \ \ (see ch.3) \ \ \ \ \ 
  $ v \ = \ 
  \Omega_{x}^{2} \oplus \Omega_{y}^{2} \oplus \Omega_{z}^{2} \oplus \Omega_{w}^{2}$ 
  with 
  
  \beq
  \omega_{x}^{2} \ = \ R_{x}^{2}R_{y}^{2}\ + \ R_{z}^{2}R_{w}^{2} \ + \ R_{y}^{2}R_{w}^{2}
  \eeq
  
  By cyclic permutation of the indices one obtains the other components as well.
  %\beqa
  %\omega_{x}^{2} \ = \ R_{x}^{2}R_{y}^{2}\ + \ R_{z}^{2}R_{w}^{2} \ + \ R_{y}^{2}R_{w}^{2} \nonumber \\
  %\omega_{y}^{2} \ = \ R_{x}^{2}R_{z}^{2}\ + \ R_{x}^{2}R_{w}^{2} \ + \ R_{y}^{2}R_{w}^{2} \nonumber \\
  %\omega_{z}^{2} \ = \ R_{x}^{2}R_{y}^{2}\ + \ R_{x}^{2}R_{w}^{2} \ + \ R_{y}^{2}R_{w}^{2} \nonumber \\
  %\omega_{w}^{2} \ = \ R_{x}^{2}R_{y}^{2}\ + \ R_{x}^{2}R_{z}^{2} \ + \ R_{y}^{2}R_{z}^{2} 
  %\eeqa
  
  These relations are identical to the minimization conditions $(4.45)$ since $ l_{i} =
  \omega_{i} R_{i}^{2}, \ \ i \ = \ x,y,z,w $ . As a result given the constants of motion $l_{x},
  l_{y}, l_{z}, l_{w}$ ,the R's are determined. 
  The stability of the spinnining $S^{3}$-brane solutions has been shown only for the radial modes.
  For the symmetric case , ( all l's, R's are equal ), we conjecture that we have full stability i.e. 
  by including perturbations of general $S^{3}$ multipole-vibrational modes.

  It is possible to choose the dimension of the ansatz $d\ = \ q_{1} + q_{2} + q_{3} + q_{4} < D-2, D $ 
  the $ p =3 $ critical dimension, i.e. $D=6,8 $ 
  and for the rest $D-2-d$ we select constant values for the coordinates
  $ X^{i}, i=D-2-d, D-1-d, \ldots, D-2 $. If $ D-2-d = 3 $  our physical space , then we have $S^{3}$
  -particles with Kaluza-Klein charges- ( internal angular momenta , as is also the case with $S^{2}$).
  The QM of the rotational modes plus quadratic vibrational ones can be carried out by using 
 only algebraic functions of 
  $ SO(q_{i})$  Casimirs.

\newpage
                                                                                                                                                                                                                                           
\section{Toroidal P-Brane Tops ($T^{2}$, $T^{3}$) on $ C^{k} \times T^{m}$} 
\subsection{$T^{2}$  Spinning Tops}

In this chapter we propose some new spinning toroidal p-brane solutions with some of the
higher dimensions compactified in Toroidal spaces.
 Double dimensional reduction of the $p=2$ Toroidal 
Supermembrane leads to type IIA string theory.
With the addition of an $S^{1}$ compactification followed by T-duality a connection is made
 with Type IIB string Theory . In order to proceed we choose $ d < D-2 $ dimensions 
to be an even number $ d=2k $. We collect the coordinates $ X^{1},X^{2}, \ldots, X^{2k-1}, X^{2k}$ into complex pairs ,

\beq
Z^{i} \ = \ X^{2i-1}\ + i \ \ X^{2i} \ \ ,\ \ \ \ \ \ i=1,\ldots, d/2
\eeq
We identify the rest ones $ D-2-d = m $  as  $Y^{a}$ with $ a=1,\ldots,m$. The Hamiltonian can be identified from ch.2 to be 

\beq
 H \ \ = \ \ \frac{T_{p}}{2} \ \ \int \ d^{p}\xi \ [ \dot{X}^{{i}^{2}} \ + 
 \ det g_{\alpha\beta } ]
\eeq 
where $ g_{\alpha\beta} \ = \ \partial_{\alpha}X^{i}\partial_{\beta}X^{i}, \ \ \ \ \
( \alpha,\beta = 1,\ldots,p) $  is the induced metric. 
The connection with the Nambu Poisson bracket is established through the identity : 
\beq
det \ g_{\alpha\beta} =  \frac{1}{p!} \ \epsilon _{\alpha_{1}\cdots \alpha_{p}}  
 \epsilon _{\beta_{1}\cdots \beta_{p}} \ g_{\alpha_{1},\beta_{1}}\cdots 
 g_{\alpha_{p},\beta_{p}} \ \ , \ \ \  \alpha_{1}(\beta_{1}),\ldots , \alpha_{p}(\beta_{p})= 1, \ldots , p 
 \eeq
for the case $p=2$ by taking into account the pairing eq.$(5.1)$ we find 

\beq
g_{\alpha\beta} \ = \ \frac{1}{2} \left( \partial_{\alpha}Z^{i} \partial_{\beta}
\bar{Z}^{i} \ + \ \partial_{\alpha}\bar{Z}^{i} \partial_{\beta} Z^{i} \right) \ + 
\ \partial_{\alpha} Y^{a}\partial_{\beta}Y^{a} \ \ , \ \ i=1,\ldots , k \ \ \ 
 a= 1, \ldots , m 
 \eeq
By applying rel.(5.3) to the case of $p=2$ Torus $T^{2}$ the Hamiltonian becomes 

\beqa
H \ &=& \ \frac{T_{p}}{2} \int d^{2}\sigma \ \ [ |\dot{Z}^{i}|^{2} \ + \ |\dot{Y}^{a}|^{2} \ + \ \frac{1}{4} 
| \{ Z^{i}, Z^{j} \} |^{2} \ +   
 \ \frac{1}{4} | \{ Z^{i},\bar{Z}^{j} \} |^{2} \ + \nonumber\\ & & 
| \{ Z^{i}, Y^{a} \} |^{2}\ + \ \frac{1}{2} | \{ Y^{a}, 
Y^{b} \} |^{2} ] \ \\ i,j &=& 1,\ldots , k \ \ \ \ \ \ a,b = 1,\ldots , m \ \ \ \ \ \  
\vec{\sigma}= ( \sigma_{1},\sigma_{2} ) \in ( 0 , 2\pi)^{2} \nonumber 
\eeqa
The constraints become :

\beq
\{ \dot{Z}^{i} , \bar{Z}^{i} \} \ + \ c.c. \ + \ \{ \dot{Y}^{a}, Y^{a} \} \ = \ 0
\eeq
 The eqs of motion for the Hamiltonian (5.5) are:
 
 \beqa
 \ddot{Z}^{i} \ &=& \ \frac{1}{2} \ \{ \{ Z^{i} , Z^{j} \} , \bar{Z}^{j} \} \ + 
 \ \frac{1}{2} \{ \{ Z^{i} , \bar{Z}^{j} \} , Z^{j} \} \ + \ \frac{1}{2} \ \{ \{ Z^{i} , Y^{a} \} , 
 Y^{a} \}  \nonumber\\  
 \ddot{Y}^{a} \ &=& \ \frac{1}{2} \ \{ \{ Y^{a} , Z^{i} \} , \bar{Z}^{i} \} \ + 
 \ \frac{1}{2} \{ \{ Y^{a} , \bar{Z}^{i} \} , Z^{i} \} \ + 
 \ \frac{1}{2} \ \{ \{ Y^{a} , Y^{b} \} , Y^{b} \}  \\
 i,j &=& 1, \ldots , k \ \ \ \ \ \ a,b= 1, \ldots , m  \nonumber
 \eeqa  
Before we proceed with the factorization ansatz let us demonstrate that the Hamiltonian
(5.5) along with the  eqs.(5.7) with a suitable dimensional reduction (double or multiple) describe LC gauge fixed closed string theory ( the Bosonic part). 
Choose all the $Y^{a}$ compactified on a torus $ T^{m}$ , $ a=1,\ldots,m$ 
with radii $R_{a}$. 

\beq
Y^{a} \ = \ R_{a} \ \cdot \ \vec{m}_{a} \cdot \vec{\xi} \ + \  
\frac{2 \pi k^{a}}{R_{a}} \cdot t 
\eeq 

$ \vec{m}_{a}=(m^{1}_{a}, m^{2}_{a}) \in Z^{2}$  the windings and  $ k^{a} $ the KK integer
momenta. 
We also assume that all the $Z^{i}, i=1,\ldots,k $ depend only on $\xi_{1}$. 
For the reduced Hamiltonian we get 

\beq
H_{red} \ = \ \pi T_{2} \ \int \ d\xi{1} \ [ \ |\dot{Z}^{i}|^{2} \ + \ k 
|\partial_{\sigma_{1}}Z^{j} |^{2} ] 
\eeq
with $ k = \sum_{a} R_{a}^{2}(m_{a}^{1})^{2} $ . By rescaling the time
$ t = \frac{1}{\sqrt{k}}\tau $ and by calling $ \xi_{1} = \xi $ we obtain

\beq
H_{string} \ = \ \frac{T_{1}}{2} \ \int_{0}^{2\pi} \ d\xi_{1} \ 
[ \ |\partial_{\tau} Z^{i}|^{2} \ + \ |\partial_{\xi} Z^{i}|^{2} ]
\eeq
where $ T_{1}= 2\pi k T_{2}$.
We will consider special embeddings of the $T^{2}$ in $ C^{k} \times T^{m}$ , 
toroidally compactified. 

\beqa
Z^{i} \ &=& \ \zeta^{i}(t) \ e^{i\vec{n}_{i}\cdot \vec{\xi}} \ \ , \ \ \ \ 
\ \ \ \ \ \ \ \ \ \ \ \ \ \ \ \ \ \ i =  1,\ldots ,k 
\nonumber \\ Y^{a}\ &=& \ R_{a} \cdot \vec{m}_{a}\cdot\vec{\xi} + 
\frac{2\pi k^{a}}{R_{a}} \cdot t \ \ , \ \ \ \ \ \ a = 1,\ldots, m  
\eeqa
$R_{a}$ are the radii of $T^{m}$ and $ \vec{m}_{a}=(m^{1}_{a}, m^{2}_{a}) \in Z^{2}$ are the winding numbers and $k^{a} $
the KK momenta. It is trivial to see that the eqs. of motion for $Y^{a}$ as well as the constraints are automatically satisfied. 
As for the Hamiltonian we find 

\beq
H = 2 \pi^{2} T_{2}  \left[ \ \sum_{i} \left( |\dot{\zeta}^{i}|^{2} + k_{i}  
|\zeta^{i}|^{2} \right)  +  \frac{1}{2}  \sum_{i,j} \ \nu_{ij} \ |\zeta^{i}|^{2}
|\zeta^{j}|^{2} \  \right ]  
\ , \ \ \ \ \  i= 1,\ldots , k 
\eeq
where 
\beq
 \nu_{ij}= ( \vec{n}_{i} \times \vec{n}_{j})^{2} \ \ \ \ \  ,  \ \ \ \ \
 k_{i}= \sum_{a}\ \  R_{a}^{2} \ (\vec{m}_{a} \times \vec{n}_{i})^{2}
 \eeq
 and $ ( \vec{n} \times \vec{m} ) \ = \ n_{1}m_{2} - n_{2}m_{1} $.
 The eqs. of motion for the $ \zeta^{i} $ are : 

 \beq
 \ddot{\zeta}^{i} \ = \ - \ \zeta^{i} \ \left( k_{i} + \sum_{j} \nu_{ij} |\zeta_{j}|^{2}
 \right) \ , \ \ \ \ \ \ i=1,\ldots , k
 \eeq
We observe that if the range of $ i-1 , \ldots , k $ is partitioned into say three
groups  $q_{1},q_{2},q_{3}$ of non-negative integers , 
with $ q_{1} + q_{2} +q_{3} = k $  and moreover 
$q_{1}$ of $ \vec{n}_{i}$ s are equal, say $ \vec{n}_{1}$ , $ q_{2}$ are equal , 
say $\vec{n}_{2}$ and the same for $q_{3}, \vec{n}_{3}$ the matrix 
$ k \times k \ \ \nu_{ij}$ has a special structure and there exist only three matrix elements which we call 
$ \nu_{12}=(\vec{n}_{1} \times \vec{n}_{2})^{2}, \nu_{23}=(\vec{n}_{2} \times \vec{n}_{3})^{2}$ as well as 
$ \nu_{31} = ( \vec{n}_{3} \times \vec{n}_{1})^{2}$. 
Furthermore we call $ \vec{w}_{1} = (\zeta^{1},\zeta^{2}, \ldots , \zeta^{q_{1}})$ , 
$ \vec{w}_{2} = ( \zeta^{q_{1}+1}, \ldots , \zeta^{q_{1}+q_{2}})$ , 
$ \vec{w}_{3} = ( \zeta^{q_{1}+q_{2}+1}, \ldots , \zeta^{k})$ the three $q_{1},q_{2},q_{3}$ 
dimensional complex vectors. Then the eqs. of motion become 

\beqa
\ddot{\vec{w}}_{1} \ &=& \ - \vec{w}_{1} \ ( k_{1}+ \nu_{12} |w_{2}|^{2} \ + \ \nu_{13} 
|w_{3}|^{2}) \nonumber \\
 k_{i} &=& \sum_{a=1}^{m} \ R_{\alpha}^{2}(\vec{m}_{\alpha} \times
\vec{n}_{i})^{2} \ \ \ \ , \ \ \ \  i=1,1,2,3 
\eeqa
Similarly for $ w_{2}, w_{3} $.

%\beqa
%\ddot{\vec{w}}_{1} \ &=& \ - \vec{w}_{1} \ ( k_{1}+ \nu_{12} |w_{2}|^{2} \ + \ \nu_{13} 
%|w_{3}|^{2}) \nonumber\\  
% \ddot{\vec{w}}_{2} \ &=& \ - \vec{w}_{2} \ ( k_{2}+ \nu_{12} |w_{1}|^{2} \ + \ \nu_{23} 
%|w_{3}|^{2}) \nonumber\\
% \ddot{\vec{w}}_{3} \ &=& \ - \vec{w}_{3} \ ( k_{3}+ \nu_{13} |w_{1}|^{2} \ + \ \nu_{23} 
%|w_{2}|^{2}) \\
% k_{i} &=& \sum_{a=1}^{m} \ R_{\alpha}^{2}(\vec{m}_{\alpha} \times
%\vec{n}_{i})^{2} \ \ \ \ , \ \ \ \  i=1,1,2,3 \nonumber
%\eeqa
with 

\beq
|\vec{w}_{1}|^{2} = \sum_{i=1}^{q_{1}} |\zeta^{i}|^{2} ,\ \ \ \ \ \ \ \ \ |\vec{w}_{2}|^{2} = 
\sum_{i=q_{1}+1}^{q_{1}+q_{2}}|\zeta^{i}|^{2},\ \ \ \ \ \ \ \ \  |\vec{w}_{3}|^{2} = 
\sum_{i=q_{1}+q_{2}+1}^{q_{1}+q_{2}+q_{3}} |\zeta^{i}|^{2}
\eeq
The Hamiltonian (5.13) now becomes 
\beq
H \ = \  2 \pi T_{2} \left[ \sum_{i=1}^{3} \ |\dot{\vec{w}}_{i}|^{2} \ + \ k_{i} |\vec{w}_{i}|^{2}
+ \nu_{12} |\vec{w}_{1}|^{2} |\vec{w}_{2}|^{2} + \nu_{23} |\vec{w}_{2}|^{2}|\vec{w}_{3}|^{2} + 
\nu_{13}|\vec{w}_{1}|^{2}
|\vec{w}_{3}|^{2} \right]
\eeq
We observe that the initial $SO(2k)$ space-rotational invariance
of the system is broken down to $ U(q_{1}) \times U(q_{2}) \times U(q_{3})$
symmetry. Note also that because of the cross product term $ \nu_{ij} $ there is a modular 
invariance $ SL(2,Z)$ which preserves $\nu_{ij}$. The new terms $ k_{i} |w_{i}|^{2}$ are harmonic
terms which are induced by the interactions of the windings $\vec{m}_{\alpha}$ with the 
$ e^{i \vec{n}_{i}\cdot \vec{\sigma}} $ dependence of the ansatz. 

The conserved "complex" angular momenta for every factor of 
$ U(q_{1}) \times U(q_{2}) \times U(q_{3})$ , call it generically U(n), are determined 
from the Hamiltonian (5.14) and Noether's theorem. The generators of U(n) are 
$n \times n $ hermitian matrices of three types. Firstly $ \frac{n (n-1)}{2} \ \ \ \ \ \ 
\ \ T^{2(ij)}$
Hermitian matrices with elements -i and i in entries  (ij) and (ji) respectively 
with zero everywhere else. Secondly there exist  
$ \frac{n (n-1)}{2} \ \ \ \ \ \ \ \ T^{1(ij)}$ Hermitian matrices  with 1 in both (ij) and 
(ji) positions with zero everywhere else and lastly n $ T^{3(ii)} $ with element 1 
in positions  (ii) and zero otherwise. For these three generators we find the conserved
angular momenta

\beqa
T_{1}^{(ij)} \ &=& \ \frac{1}{2} \ \left( z^{i} \dot{\bar{z}}^{j} \ + \ \bar{z}^{i} 
\dot{z}^{j} \right) - \frac{1}{2} \left( z^{j} \dot{\bar{z}}^{i} \ + \ \bar{z}^{j} 
\dot{z}^{i} \right) \ , \ \ \ \ \ i>j=1,\ldots , n \nonumber \\
T_{2}^{(ij)} \ &=& \ \frac{-i}{2} \ \left( z^{i} \dot{\bar{z}}^{j} \ - \ \bar{z}^{i} 
\dot{z}^{j} \right) - \frac{i}{2} \left( z^{j} \dot{\bar{z}}^{i} \ - \ \bar{z}^{j} 
\dot{z}^{i} \right) \ , \ \ \  i>j=1,\ldots , n \nonumber \\
T_{3}^{(ii)} \ &=& \ \frac{-i}{2} \ \left( z^{i} \dot{\bar{z}}^{i} \ - \ \bar{z}^{i} 
\dot{z}^{i} \right)  \ \ , \ \ \ \ \ \ \ \ \ \ \ \ \ \ \ \ \ \ \ \ \ \ \ \ 
\ \ \ \  i=1,\ldots , n
\eeqa 
These are real conserved quantities which can be grouped into one complex and one real as
follows: 

\beqa
T^{(ij)} \ &=& \ T_{1}^{(ij)} +  T_{2}^{(ij)} = z^{i}\dot{\bar{z}}^{j} - \bar{z}^{j} 
\dot {z}^{i} \ \ ,\ \ \ \ \ \ \ \ \ \ \ \ \ \ \ \ i>j=1,\ldots,n \nonumber\\
T_{3}^{(ii)} \ &=& \ - \frac{i}{2} \ \ T^{(ii)} 
\eeqa
By using some familiar identites we demonstrate that the Casimir element 
\beq
\sum_{i>j} \ \left[ \left( T_{1}^{(ij)} \right)^{2} \ + \ \left( T_{2}^{(ij)}\right)^{2} 
\right] \ + \ \sum_{i} \ \left( T_{3}^{(ii)}\right)^{2} \ \equiv \ \vec{T}^{2}
\eeq
is related to the generic kinetic term 
\beq
|\dot{\vec{z}}|^{2} \ = \ |\dot{z}_{1}|^{2} + \cdots + |\dot{z}_{n}|^{2} \ = \ 
\frac{ \vec{T}^{2}}{r^{2}} \ + \ \dot{r}^{2}
\eeq
where $ r^{2} = |z_{1}|^{2} + \cdots + |z_{n}|^{2} $  
 So if we call the lengths of the complex vectors $|\vec{w}^{i}|=r_{i} , \ \ i=1,2,3 $ and
the Casimirs of each factor $U(q_{i})$ \ $T_{i}^{2} = \vec{T}_{i}^{2},\ \ \ i=1,2,3$ 
the Hamiltonian can be written as 
\beq
H \ = \ 2 \pi^{2} T_{2} \left[ \ \ \sum_{i=1}^{3} \ \ \left( \dot{r}_{i}^{2} \ + 
\ \frac{ T_{i}^{2}}{r_{i}^{2}} \right) \ + \ \sum_{i=1}^{3} \ k_{i} r_{i}^{2} \ + \ 
\nu_{3} r_{1}^{2} r_{2}^{2} + \nu_{2} r_{1}^{2} r_{3}^{2} + \nu_{1} r_{2}^{2} r_{3}^{2} \right]
\eeq
with $ \nu_{1} = \nu_{23} , \nu_{2} =\nu_{3} , \nu_{3}=\nu_{12} $.

In order to obtain the Euler-Top solutions we proceed as with the spherical cases 
$(S^{2}, S^{3})$ of ch.4. The energy minimization conditions for constant radii 
$ r_{i}, i=1,2,3 $ are as follows :
\beq
T_{1}^{2} \ = \ r_{1}^{4} \ ( k_{1} \ + \ \nu_{3} r_{2}^{2} \ + \ \nu_{2}r_{3}^{2})
\eeq
Similarly for the other components.

%\beqa
%T_{1}^{2} \ &=& \ r_{1}^{4} \ ( k_{1} \ + \ \nu_{3} r_{2}^{2} \ + \ \nu_{2}r_{3}^{2}) 
%\nonumber\\ T_{2}^{2} \ &=& \ r_{1}^{4} \ ( k_{1} \ + \ \nu_{3} r_{2}^{2} \ + 
%\ \nu_{2}r_{3}^{2}) \nonumber\\ 
%T_{3}^{2} \ &=& \ r_{1}^{4} \ ( k_{1} \ + \ \nu_{3} r_{2}^{2} \ + \ \nu_{2}r_{3}^{2})
%\eeqa
We observe that the permutation symmetry $ r_{1} \leftrightarrow r_{2} \leftrightarrow 
r_{3} $ is broken unless we have the special point $k_{1}=k_{2}=k_{3}=k,\nu_{1}=\nu_{2}=\nu_{3}
=\nu$ . If we choose $ \vec{n}_{1}+\vec{n}_{2}+\vec{n}_{3}=0$ (special embeddings) 
 then we guarrantee that $\nu_{1}=\nu_{2}=\nu_{3}=\nu$ . We proceed to solve (5.23) for the
 special point 
 
 \beq
 T_{1}^{2} \ \ = \ \ r_{1}^{4} \ \left( k \ + \ \nu ( r_{2}^{2}+r_{3}^{2}) \right)
 \eeq
 Similarly for the other components.
 %\beqa
 %T_{1}^{2} \ \ &=& \ \ r_{1}^{4} \ \left( k \ + \ \nu ( r_{2}^{2}+r_{3}^{2}) \right) \nonumber\\
 %T_{2}^{2} \ \ &=& \ \ r_{2}^{4} \ \left( k \ + \ \nu ( r_{1}^{2}+r_{3}^{2}) \right)\nonumber\\
 %T_{3}^{2} \ \ &=& \ \ r_{3}^{4} \ \left( k \ + \ \nu ( r_{1}^{2}+r_{2}^{2}) \right)
 %\eeqa
 We observe that the difference with the previous $S^{2}$ case lies in the harmonic term k.
 For the completely symmetric case (symmetric toroidal 2-brane Top) 
 $ T_{1}^{2} =T_{2}^{2} =T_{3}^{2}=T_{s}^{2} $ and 
 $ r_{1}^{2} =r_{1}^{2} =r_{1}^{2}=r_{s}^{2} $ we find 
 \beq
 T_{s}^{2} \ \ = \ \ r_{s}^{4} \ \left( k \ + \ 2 \ \nu \ r_{s}^{2} \right)
 \eeq
 while for the axially symmetric case $ r_{1}=r_{2}=r \ \ ,\ \ T_{1}=T_{2}=T$ we obtain 
 
 \beqa
 T^{2} \ &=& \ r^{4} \left( k \ + \ \nu \ ( r^{2} \ + \ r_{3}^{2}) \right) \nonumber\\
 T_{3}^{2} \ &=& \ r_{3}^{4} \ \left( k \ + \ 2 \ \nu  r_{3}r^{2}\right)
 \eeqa
 For the symmetric case it is possible to get an analytic expression for the solution
 which follows from the careful analysis of the cubic equation (5.25) . 
 We define two ratios 
 \beq
 \rho_{k} \ = \ \left( \frac{k}{6\nu}\right)^{3} ,\ \ \ \rho_{T} \ = \ \frac{T^{2}}{4\nu}
 \eeq
For $ \rho_{T} > 2 \rho_{k} $ the equilibrium value of the radius of the torus 
which corresponds to  the balancing
out of the attractive tension against the repulsive algular kinetic energy is found to 
be

\beq
r_{s}^{2} \ = \ \left[ \rho_{T} - \rho_{k} + \sqrt{\rho_{T}(\rho_{T} - 2\rho_{k})} 
\right]^{1/3} + \left[ \rho_{T} - \rho_{k} - \sqrt{\rho_{T}(\rho_{T}-2\rho_{k})}
\right]^{1/3} - \rho_{k}^{1/3}
\eeq
For $\rho_{T} < 2 \rho_{k}$ combinations with the third root of unity 
$ e^{2\pi i/3}$ give the result. The eq.(5.25) has always one largest positive root.
The energy of the solution is 
\beq
E_{s} \ = \ 8 \ \pi^{2} \ T_{2} \ \nu \ \left( r_{s}^{4} \ + \ 4 \ \rho_{k}^{1/3} \ 
r_{s}^{2} \right)
\eeq
For large angular 
momenta $ \rho_{T} \gg \rho_{k}, \rho_{T} \rightarrow \infty $ the radius $  r_{s}^{2} $
behaves like
\beq
r_{s}^{2} \ \sim \ \left(\frac{T^{2}}{4 \nu }\right)^{1/3}
\eeq
while the energy scales like
\beq
E_{s} \ \sim \ \left( \nu \ T \ \right)^{4/3}
\eeq
We have an identical power law behaviour with the $S^{2}$ case ( apart from the factor 
$ \nu=(\vec{n}_{1} \times \vec{n}_{2})^{2}$ see rel.(5.13).
The axially symmetric case (5.26) is algebraically not tractable apart from 
some special points in the space of parameters $\nu,k, T^{2},T^{3}$.
We now proceed to discuss the time dependence of the complex vectors 
$ \vec{w}_{i}, i=1,2,3$.  
\beq
\vec{w}_{i} \ = \ e^{i \Omega_{i} t} \ \ \vec{w}_{i}(t=0)
\eeq
In general $\Omega_{i}$ is a linear combination of the $ T_{1},T_{2}, T_{3}$ hermitian 
matrices discussed previously. By using $U_{q_{i}}$ transformations we can bring 
$\vec{w}_{i}(t=0)$,$\dot{\vec{w}}_{i}(t=0)$ in the ($z_{1},z_{2}$) complex plane and the 
$\Omega_{i}$ have the form of an $SU(2)$ hermitian matrix. In the simplest case 
 of a diagonal U(1) matrices we get angular velocities $\omega_{i}$ which satisfy the 
(5.15) eqs. of motion 
\beq
\omega_{1}^{2} \ = \ k_{1} \ + \ v_{3} r_{2}^{2} \ + \ v_{2} r_{3}^{2}
\eeq
Similarly for $\omega_{2}, \omega_{3}$.
%\beqa
%\omega_{1}^{2} \ &=& \ k_{1} \ + \ v_{3} r_{2}^{2} \ + \ v_{2} r_{3}^{2} \nonumber\\
%\omega_{2}^{2} \ &=& \ k_{2} \ + \ v_{3} r_{1}^{2} \ + \ v_{1} r_{3}^{2} \nonumber\\
%\omega_{3}^{2} \ &=& \ k_{3} \ + \ v_{2} r_{1}^{2} \ + \ v_{1} r_{2}^{2} 
%\eeqa
We can check from (5.21) 

\beq
T_{i}^{2} \ = \ \omega_{i}^{2} \cdot r_{i}^{4} \ \ , \ \ \ \ \ \ i=1,2,3
\eeq
and so the minimization conditions are identical with the eqs. of motion (5.15).

\subsection{The Three Dimensional Spinning Torus \ \ $ T^{3}$}

Our last but not least example of spinning p-brane is the spinning $T^{3}$ torus 
 $(p=3)$. The example is the richest one which exhibits unitary group symmetries 
 $ \prod_{i=1}^{4} U(q_{i})$ as well as a larger modular group symmetry $SL(3,Z)$. 
 Moreover , it leads to the $p=2$ (membrane) case by double dimensional reduction. 
This is an extension of the reduction of the membrane $(p=2)$ to the string case 
 $(p=1)$. 

We start again from the basic Hamiltonian
\beq
H \ = \ \frac{T_{p}}{2} \ \ \int \ d^{3}\xi \ \ \left[ \dot{X}^{i^{2}} \ + \ 
det \ \ ( \partial_{\alpha} X^{i} \partial_{\beta}X^{i})  \ \right]
\eeq
and the constraints 

\beq
\left\{  \dot{X}^{i} , X^{i} \right\}_{\alpha ,\beta} \ = \ 0 \ \ \ \ 
\alpha \neq \beta = 1,2,3 \ , \ \ \ \ i=1 ,\ldots , D-2
\eeq
The volume preserving diffeomorphisms contain also global translations 
$ P_{\alpha} ,\ \ \alpha=1,2,3 $ along cycles at $T^{3}$ which are not 
connected to the identity. The connected subgroup is generated  polynomially
by the Nambu-Bracket algebra. 

\beq
\left\{ e_{\vec{n}_{1}} \ , \ e_{\vec{n}_{2}} \ , \  e_{\vec{n}_{3}} \right\} \  =  \ i^{3}
 \ det \left[ \vec{n}_{1} , \vec{n}_{2} , \vec{n}_{3} \right ] \ \cdot \ 
e_{\vec{n}_{1}+\vec{n}_{2}+
\vec{n}_{3}}
\eeq
where the basic functions $ e_{\vec{n}} $ are : 
\beq
e_{\vec{n}} \ = \ e^{i \vec{n} \cdot \vec{\xi}} \ \ \ \ , \ \ \ \  
\vec{\xi} \in [ 0 , 2\pi ]^{3} \ \ , \ \ \vec{n} = ( n^{1}, \ n^{2}, \ n^{3} ) \in 
 Z^{3}
\eeq 
and

\beq
det \left[ \vec{n}_{1} , \vec{n}_{2} , \vec{n}_{3} \right ] = \epsilon_{\alpha\beta\gamma} 
\  n_{1}^{\alpha} n_{2}^{\beta} n_{3}^{\gamma} 
\eeq
The automorphism group of (5.38) contains the SL(3,Z) modular group which leaves 
the structutre constants invariant $ \vec{n} \rightarrow  A \vec{n} $
, $ A \in $ SL(3,Z) ,$ A =(A_{ij})$ integer matrix with $ det A=1 $

In order to proceed with our ansatz we separate the D-2 target coordinates 
$ X^{i}, i= 1, \ldots, D-2 $ into two groups. Firstly we pair $ X^{1},X^{2},\ldots, X^{2k}$ 
into complex ones  $ 2k < D-2 $ 
\beq
Z^{l} = X^{2l-1} + i X^{2l} \ \ , \ \ \ \ \ \ l=1,\ldots , k 
\eeq
 and the rest $ D-2-2k \equiv  m ,\ \ \  Y^{\alpha}, \alpha=1,\ldots,m $. The determinant 
 $ det g_{\alpha\beta} $ of the
induced metric : 
\beq
g_{\alpha\beta} = \frac{1}{2} \ \partial_{\alpha}Z^{l}\partial_{\beta}
\bar{Z}^{l} + \frac{1}{2} \partial_{\alpha}\bar{Z}^{l}\partial_{\beta}
Z^{l} + \partial_{\alpha}Y^{a}\partial_{\beta}
Y^{a} \ \ \ , \ \ \ \ l=1, \ldots ,k \ \ \ ,a = 1,\ldots, m \ \ ,\ \ 
\alpha,\beta=1,2,3 
\eeq  
can be calculated. 
We derive the Hamiltonian in terms of $Z^{l}$, $ Y^{a}$ s :

\beqa
H \  &=&  \  \frac{T_{3}}{2} \int d^{3} \xi  \ ( \  | \dot{Z}^{i} |^{2} \  +  \  
  |\dot{Y}^{a} |^{2} \  +  \ \frac{1}{24} \ |\{ Z^{i} , Z^{j} , Z^{k} \} |^{2} \ +\nonumber \\ 
 & & \frac{1}{8} \  | \{ Z^{i} , Z^{j} , \bar{Z}^{k} \} |^{2} \ +  
 \frac{1}{4}  \ | \{ Z^{i} , Z^{j} , Y^{a} \} |^{2}  \ +    
  \frac{1}{4}  \ |\{ Z^{i} , \bar{Z}^{j} , Y^{a} \} |^{2} \  + \nonumber  \\  
 & &\frac{1}{2}  \ | \{ Z^{i} , Y^{a} , Y^{b} \} |^{2} \ + \  
\frac{1}{3!}  \ |\{ Y^{a} , Y^{b} , Y^{c} \} |^{2} \ )   
\eeqa
with $ a,b,c= 1,\ldots, m $  and the constraints :

\beq
\left \{ \dot{Z}^{i} , \bar{Z}^{i} \right \} + c\cdot c + \left \{ \dot{Y}^{a}, 
Y^{a} \right \}_{a,b} = 0 \ , \ \ \ a,b = 1,2,3
\eeq
We notice here that upon double dimensional reduction that is, by compactification
on a circle and assuming that $Z^{i}\ \ , \ \ i=1 , \ldots,k $ depend only on 
$ \xi^{1},\xi^{2} $ and not on $\xi^{3}$ , we can get from the Hamiltonian of $p=3$
(5.42) toroidal branes of the $p=2$ type. Indeed the above asumption leads 
to ( constant terms are neglected) : 

\beqa
H \ &=& \ \frac{T_{3}}{2} \ \ 2 \pi \ \ \int d^{2}\xi \ \ [ \ |\dot{Z}^{i}|^{2} \ + \ 
\frac{1}{4} \ \nu \ ( \ | \{ Z^{i} , Z^{j} \} |^{2} \ + \ | \{ Z^{i}, \bar{Z}^{j} \} |^{2})
\ + \nonumber \\ & &  \frac{1}{2} |L_{a,b} Z^{i} |^{2} \ \ ] 
\ \ \ \ , \ \ \ \ i,j=1,\ldots , k 
\ \ \ a,b = 1 ,\ldots , m 
\eeqa
with 

\beqa
L_{a,b} \ &=& \ R^{a}R^{b} \ \left[ ( m^{a}_{2}  m^{b}_{3} \ - \ 
m_{3}^{a} \ m^{b}_{2}) \partial_{1} \ \ + \ \ ( m^{a}_{3}  m^{b}_{1} \ - \ m_{1}^{a} \ m^{b}_{3}) \partial_{2} \right]  \nonumber \\ 
\nu &=& \sum_{a} ( R^{a} m^{a}_{3})^{2} \ \ , \ \ \ \ \ \ \ \ \ \
a, b =1, \ldots , m
\eeqa
With appropriate diagonalization and rescaling of the operator $ L_{a,b} $ we
can arrive at normal form of the harmonic term $ | L_{a,b} \tau^{i} |^{2} $ 
and derive eqs. of motion for $T^{2}$. The compactified target coordinates induce, constant , 
harmonic  and unharmonic terms respectively on the Hamiltonian. The constant term 
corresponds to the KK kinetic energy as well as the winding energy $ \sum_{a,b,c}
( R^{a} R^{b} R^{c})^{2} det^{2}( \vec{m}^{a} 
\vec{m}^{b}\vec{m}^{c}) $.
The reduced Hamiltonian without the constant term is as follows (summation over
the indices is implied):

\beqa
H \ &=& \ \frac{T_{3}}{2} \ (2\pi)^{3} \ [ \ |\dot{\zeta}^{i}|^{2} \ + \ \frac{1}{6} \ 
det^{2}(\vec{n}_{i}\vec{n}_{j}\vec{n}_{k}) |\zeta^{i}|^{2}|\zeta^{j}|^{2}|\zeta^{k}|^{2}
 \nonumber \\ 
& & +\frac{1}{2} \ |\zeta^{i}|^{2}|\zeta^{j}|^{2} R^{a^{2}} \ det^{2} ( \vec{n}_{i}, 
\vec{n}_{j}, \vec{m}^{a} ) \  \nonumber\\ 
& &  + \frac{1}{2} \ |\zeta^{i}|^{2} \ \ R^{a^{2}}
R^{b^{2}} det^{2}( \vec{n}_{i} , \vec{m}^{a}, \vec{m}^{b}) ]
\eeqa 
and the $\zeta^{i}$ complex scale factors satisfy the eqs. :

\beq
\ddot{\zeta}^{i} \ = \ -\frac{1}{2} \ \zeta^{i} \ \left[ \sum_{i,l\neq i} |\zeta^{j}|^{2}
|\zeta^{k}|^{2} \lambda_{ijk} \ + \ 2 \sum_{j\neq i} |\zeta^{j}|^{2}\nu_{ij} \ + \ 
\dot{k}_{i} \ \right] 
\eeq
with
\beqa
\lambda_{ijl} \ &=& \ det^{2}(\vec{n}_{i},\vec{n}_{j},\vec{n}_{l}) \ , \ \ \ \ 
i \neq j \neq l =1,\ldots , k \nonumber \\
\nu_{ij} \ &=& \ \sum_{\alpha} \ R^{\alpha^{2}} \ det^{2}(\vec{n}_{i},\vec{n}_{j},
\vec{m}^{\alpha}) \nonumber \\
k_{i} \ &=& \ \sum_{\alpha\neq\beta} R^{\alpha^{2}} R^{\beta^{2}} det^{2}
(\vec{n}_{i},\vec{m}_{\alpha},\vec{m}^{\beta})
\eeqa
We now have the options to either use the ansatz of many $U(1)$ s (ref.\cite{AFP}) 
\beq
\zeta^{i} \ = \ R^{i} \ e^{i\omega_{i} t} \ \ \ \ , \ \ \ \ \ i=1,\ldots , k 
\eeq
or to form 4-complex vectors of $ q_{j} , j=1,2,3,4 $ components, of $q_{j}$ $ \zeta^{i}$ s
$j=1,2,3,4 $ which possess only four different $\vec{n}_{i}$ say $q_{1}\ \ \ \vec{n}_{1}$'s,
 $q_{2}\ \ \ \vec{n}_{2}$'s  and we make the ansatz :

\beqa
\vec{w}_{1} \ &=& \ ( \ \zeta^{1},\zeta^{2},\ldots , \zeta^{q_{1}} ) \ e^{i\vec{n}_{1}\cdot\vec{\xi}}
 \nonumber \\
\vec{w}_{2} \ &=& \ ( \ \zeta^{q_{1}+1},\ldots , \zeta^{q_{1}+q_{2}} ) \ 
 e^{i\vec{n}_{2}\cdot\vec{\xi}} \nonumber \\
\vec{w}_{3} \ &=& \ ( \ \zeta^{q_{1}+q_{2}+1},\ldots , \zeta^{q_{1}+q_{2}+q_{3}} ) 
\ e^{i\vec{n}_{3}\cdot\vec{\xi}}  \\
\vec{w}_{4} \ &=& \ ( \ \zeta^{q_{1}+q_{2}+q_{3}+1},\ldots , \zeta^{k} ) 
\ e^{i\vec{n}_{4}\cdot\vec{\xi}} \nonumber \\
 & &   q_{1}+q_{2}+q_{3}+q_{4} = k \nonumber
\eeqa 
The resulting Hamiltonian is: 
\beqa
H \ &=& \ \frac{T_{3}}{2} ( 2 \pi )^{3} \ \ [ \ \sum_{i=1}^{4} \ \left( \ \dot{r}^{2}_{i} \ + \ 
\frac{T_{i}^{2}}{r_{i}^{2}} \right) \ + \ \frac{1}{6} \ \sum_{i\neq j \neq k=1}^{4} \ 
\lambda_{ijk} \ r_{i}^{2} r_{j}^{2} r_{k}^{2} \ \nonumber \\
 & &  + \ \frac{1}{2} \ \sum_{i\neq j=1}^{4} \ 
\nu_{ij} \ r_{i}^{2} \ r_{j}^{2} \ + \ \frac{1}{2} \ \sum_{i=1}^{4} k_{i} \ r_{i}^{2}
\eeqa
where $ r_{i}^{2} \ = \ | \vec{w}_{i} |^{2} $ , and 
\beq
 | \dot{\vec{w}}_{i} |^{2} \ = \ \dot{r}_{i}^{2} \ + \ \frac{T_{i}^{2}}{r_{i}^{2}} \ \ \ \ ,
 \ \ \ i=1,2,3,4 
\eeq
with $ T_{i}^{2}$ being the $ U(q_{i}) ,\ \  i=1,2,3,4 $ Casimirs . 
The time dependence of the ansatz is given by :

\beq
\vec{w}_{i}(t) \ = \ e^{i \Omega_{i} \cdot t} \ \vec{w}_{i}(o) \ \ \ \ \ , \ \ \ \ i=1,2,3,4
\eeq 
with $ \Omega_{i}$ being the generators of $ U(q_{i})$ . By diagonalizing the 
$ \Omega_{i}$
as in the case of $ T^{2}$ in the appropriate complex planes of $\vec{w}_{i}$ s 
we get from the eqs of motion 

\beqa
\omega_{1}^{2}  \ &=& \  \frac{1}{3} \ ( \lambda_{123} \ r_{2}^{2} \ r_{3}^{2} \ + 
\ \lambda_{134} \ r_{3}^{2} \ r_{4}^{2} ) \ \nonumber \\  
& &  + \ \nu_{12} \  r_{2}^{2} \ + \ \nu_{13}  \ r_{3}^{2} \ + 
\ \nu_{14}  \ r_{4}^{2} \ + \ k_{1}
\eeqa
and cyclically for the other indices $ i=1,2,3,4$ . 

These relations correspond to nothing else but the minimization conditions for the
effective potential ( since $ T_{i}^{2} \ = \ \omega_{i}^{2} \ r_{i}^{4} $)

\beq
 V_{eff} \ = \ H \ - \ \frac{(2\pi)^{3}}{2} \ T_{3} \ \sum_{i=1}^{4} \dot{r}_{i}^{2}
\eeq
For the completely symmetric $T^{3}$ with symmetric initial conditions,
( which can be satisfied if $ \vec{n}_{1}+\vec{n}_{2}+\vec{n}_{3}+\vec{n}_{4}=0 $),
$ T_{i}=T \ , \ 
 k_{i}=k \ , \ \nu_{ij}=\nu \ , \ \lambda_{ijk}=\lambda  \ \ , \ \ 
 i \neq j \neq k=1,2,3,4 $ we obtain 
 
 \beq
 T^{2} \ \ = \ \ r^{4} \ \ \left( \lambda \ r^{4} \ \ + \ \ \nu \ r^{2} \ \ + \ \ k \right)
 \eeq
setting $ r^{2}=u$ we obtain the 4rth order polynomial equation 
\beq
T^{2} \ \ = \ \ \lambda \ u^{4} \ \ + \ \ \nu \ u^{3} \ \ + \ \ k \ u^{2} 
\eeq 
which can be solved by quadratures . Indeed there exist two real roots for u 
( one positive and one  negative) as well as a pair of complex conjugate roots , 
for positive $ T^{2}, \lambda , \nu , k $ . The energy of the configurations is 
expressed in terms only of $ \lambda, \nu, k$ through the positive root $ u_{s} $
of (5.57) ($ u_{s} = r_{s}^{2})$ 

\beq
E_{s} \ = \ \frac{T_{3}}{2} \  \ (2 \pi )^{3} \  \ \left[ \ \frac{14}{3} u_{s}^{3} \ + \ 
 15 \ \nu \ \ u_{s}^{2} \ \ + \ \ 6 \ k \ u_{s} \ \right]
\eeq
For large values of the angular momenta $ T^{2} \rightarrow  \infty $ the energy scales as :

\beq
E \ \ \sim \ \ \left( \ \lambda \ T^{2} \ \right)^{3/4}
\eeq

\section{Interpretation of the Results \ - \ Conclusions}

We have been working in this paper with the Light Cone Gauge fixed Hamiltonian of the
Nambu-Goto p-branes. The target space dimensions for various p, $ p=1,2,3,\ldots $ are
restricted by target space and k-world volume supersymmetry
in order that physical bosonic and fermionic degrees to match. 
The relevant brane scan determines these dimensions. For
$ p=1, D=3,4,6,10 $ for $p=2, D=4,5,7,11 $ , for $p=3 , D=6,8 $ and $ p=4 , D=9 $ and finally
$p=5, D=10 $. 
If one adds gauge and tensor fields on the world volume of the branes there are additional 
restrictions\cite{Berg}. In this case the p-branes are charged under the gauge groups. For compact p-branes
the total charge must be zero(Gauss-Law).

With the advent of $D_{p}$-branes \cite{Polc} it was understood 
that there are solitonic objects of type IIA-B superstring theories in $D=10$, where for
IIA theories $ p=0,2,4,6,8 $ carrying NS-NS charges and for IIB theories $ p=1,3,5,7,9 $
carrying RR charges respectively. The most intriguing 
ones are of the IIA type $p=2$ super D-membrane and of the type IIB $p=3$ selfdual one
along with the $p=5$ famous fivebrane. The D-branes apart from being the sought 
after sources of RR and NS charges they have more degrees of freedom , the various 
p-form gauge world volume fields. Although so rich in structure and so well 
studied they have infinite extent (infinite charge and energy). Their finite 
(charge and energy ) cousins (the Nambu-Goto p-branes) still escape our ability
to describe them dynamically (unless compactified  
on various compact submanifolds) due to strong string coupling. 

Our solutions are not charged but if we turn on the 11-dim flux field then the total 
charge becomes zero but with the dipole and multipole moments non-zero. 
In the latter case the equations of motion get modified.
The simplest case is the 11dim. pp-wave background with a constant flux \cite{Jabb}

In a relatively recent work J.J. Rousso et.al. \cite{Russo} studied rotating toroidal 
p-branes ( Nambu-Goto ones) and observed that they represent type IIA-B solitons. 
Essentially the argument for type IIA is that the tension $T_{2}$ of the $p=2$ membrane
compactified from $D=11$ dim. by double dimensional reduction to $D=10$ on 
a circle of radius $R_{10}$ goes like $ T_{2} \ = \ \frac{T}{g_{IIA}} $ for fixed string tension
T. Similarly one has $ T_{2} \ = \ \frac{T}{g_{IIB}} $  
for  type IIB string theory which is
compactified directly from $ D=11$ to $ D=9$ on a Torus $ T^{2} = S^{1} \times S^{1}$ 
followed by a T-duality on the second $S^{1}$ . 
In the above work the solutions are found in a covariant gauge  $ X^{o}= P \cdot \tau $ 
and there are constraints which cannot be solved except for in some special cases.

In our examples (ch. 4-5) the rotating $p=2$ solutions are given in the 
light-cone gauge  where the constraints are solved automatically by the ansatz. 
The nice arguments of J.J.Rousso et.al. for the solitonic character of the $p=2$ 
Toroidal membranes go through also in our case. Moreover we presented new results
for $S^{3}$ , $ T^{3}$ spinning $p=3$ branes .  We 
would like to call these solutions massive giant gravitons of flat spacetimes. 
Our solutions are embedable in lightcone pp-wave backgrounds with fluxes. On 
these problems we are currently at work. More general backgrounds like $G_{2}$, 
$ AdS^{7} \times S^{4} $ etc. are expected to host such solutions although 
in these cases the constraints in general cannot be solved \cite{Alish,UeYa,Yon}.

In conclusion, we have constructed new spinning $p=2 \ \ S^{2},T^{2}$ and 
$p=3 \ \ S^{3},T^{3}$ Nambu-Goto
p-branes which behave like Eulers Tops with higher rotational symmetries
$ \prod_{i}(SO(q_{i})) $ or $ \prod_{i}(SU(q_{i})) $ respectively. 
This is due to the balancing out of the 
attractive brane-tension forces in higher dimensions 
by the repulsive effect of rotation alone for
$S^{2}, S^{3}$ and in conjunction  with the induced harmonic forces arising from Toroidal
Compactifications for $T^{2}, T^{3}$. The minimization of the energy led to 
its unique scaling with the angular momenta ( and for $T^{2}, T^{3}$ from the winding ).
For the case of $T^{2}$ ( and presumably $T^{3}$ ) the energy has solitonic dependence
on type IIA, IIB string couplings. These solutions can be thought of as particle
like objects with quantum numbers  $ \prod_{i}  \  SO(q_{i})$ 
 for  $ S^{2}, S^{3} $ and $ \prod_{i} \ U(q_{i})$ for $ T^{2}, T^{3}$ . For the
case of completely symmetric configurations  $ S^{2},S^{3} $ or $ T^{2},T^{3}$
( same radii in all dimensions) the equilibrium equations can be solved
analytically by elliptic integrals as well as the angular velocities time dependence.
These configurations , however, are only excitations of the Euler Tops. 

One of the many interesting open questions is the Matrix model construction and the
corresponding Euler-Top solutions in flat or pp-wave backgrounds for p-branes. 
There are several attempts \cite{Ramg,Alish}, but we think that more 
drastic propositions like the works of H.Awata et.al in \cite{Nam} should be given more
attention\cite{Ceder}
Among possible interesting applications would be a spinning brane-world 
scenario for both $S^{3}$ 
and $ T^{3}$ solutions.

\section{Acknowledgments}
For discussions we thank A.Kehagias, A.Petkou and G. Savvidy. 
We thank C.Kokorelis for collaboration in the initial phases of the work. 
The research of both M.A. and E.F. was partially supported by E.U. grants: 
MRTN-CT-2004-512194-503369.

\end{document}